\documentclass{article}

\usepackage{arxiv}

\usepackage[utf8]{inputenc} 
\usepackage[T1]{fontenc}    
\usepackage{hyperref}       
\usepackage{url}            
\usepackage{booktabs}       
\usepackage{amsfonts}       
\usepackage{nicefrac}       
\usepackage{microtype}      
\usepackage{lipsum}
\usepackage{graphicx}
\usepackage{amsmath}
\usepackage{subcaption}
\usepackage{authblk}
\usepackage{changepage}
\usepackage{amsmath}
\usepackage{amssymb}
\usepackage{enumitem}
\usepackage[section]{placeins}

\graphicspath{{./picture/}}

\title{Model-Free Current Control of Permanent Magnet Synchronous Motors via ESO-Based Disturbance Feedforward and Data-Driven \(H_\infty\)  Residual Feedback}

\author{
  YongBo Li, Shuang Liang, HongWei Ma \and
  \texttt{School of Automation, Beijing Institute of Technology}
}

\begin{document}
\maketitle
\begin{abstract}
This paper proposes a model-free current control method for permanent magnet synchronous motors (PMSMs) based on a functional decomposition into disturbance feedforward and residual feedback. Using an ultra-local current model, motor dynamics, parameter uncertainties, cross-coupling effects, and other nonideal factors are collectively incorporated into generalized lumped disturbances. An extended state observer (ESO) is employed to estimate the lumped disturbances and compensate for them through feedforward action, thereby transforming the original PMSM current-control problem into the regulation of a simplified post-compensation residual system. Instead of directly designing a feedback controller for the original motor dynamics, a state-feedback \(H_\infty\) controller for the residual system is learned directly from operating data using off-policy integral reinforcement learning. Owing to the simplified residual dynamics, the value function and control policies can be parameterized in quadratic and linear forms, respectively, reducing the off-policy learning problem to a low-dimensional parameter-estimation problem without requiring neural-network approximation. Consequently, the proposed method requires neither prior knowledge nor online identification of PMSM electrical parameters: input-gain mismatch is incorporated into the ESO-estimated lumped dynamics, while the residual-feedback policy is obtained from operating data. Comparative simulations against deadbeat predictive current control, model-based \(H_\infty\) control, and model-free predictive current control demonstrate fast current tracking, low current distortion, and strong robustness to large variations in PMSM electrical parameters. Notably, with the learned \(H_\infty\) feedback policy kept fixed and without retraining or controller retuning, the proposed method maintains nearly unchanged current-control performance when the stator resistance, stator inductance, and permanent-magnet flux linkage are simultaneously varied to 20\% and 200\% of their nominal values.
\end{abstract}

\section{Introduction}
Permanent magnet synchronous motors (PMSMs) are widely employed in high-performance electric-drive applications, ranging from electric vehicles to robotic actuators, owing to their high efficiency, high power density, and favorable torque characteristics. Since electromagnetic torque regulation fundamentally relies on accurate stator-current control, the performance of the inner current loop directly affects the dynamic response, steady-state accuracy, and current quality of the overall drive system. Conventional current-control approaches, such as predictive current control and model-based robust control, are generally designed on the basis of the electrical model of the motor[1,2]. However, key electrical parameters such as the stator resistance, stator inductance, and permanent-magnet flux linkage may vary with temperature, magnetic saturation, and operating conditions. Such variations introduce discrepancies between the nominal model used for controller design and the actual plant, potentially degrading current-tracking performance and current quality and requiring motor-specific controller recalibration or retuning. Reducing both the dependence on accurate motor parameters and the associated tuning burden therefore remains an important issue in high-performance PMSM current control.

Model-free control offers an attractive route to reducing such model dependence. The ultra-local model introduced in model-free control represents the input-output behaviour of a plant without requiring an explicit description of its internal physical dynamics[3]. Related disturbance-estimation concepts, particularly extended-state-observer-based formulations, provide an effective means of reconstructing and compensating generalized disturbances and uncertainties[4,5]. Model-free control has subsequently been applied to PMSM drives[6], while a range of ultra-local-model-based predictive and deadbeat current-control methods have been developed to improve parameter robustness[7-9,12-15]. Among them, ESO-based schemes estimate the lumped motor dynamics and compensate for their effects without explicitly reconstructing the complete PMSM electrical model[7,9]. Parameter-free, self-commissioning, finite-time gain-estimation, and adaptive-gain approaches have also been investigated to further reduce dependence on motor parameters or input-gain information[10-12,14,15]. These developments demonstrate the effectiveness of ultra-local disturbance compensation; nevertheless, the dynamics remaining after feedforward compensation must still be appropriately regulated, and the corresponding feedback design may introduce additional gain selection or adaptation requirements.

Data-driven optimal and robust control provides a complementary approach. Policy-iteration and reinforcement-learning methods enable control policies to be obtained from system trajectories without requiring complete knowledge of the underlying plant dynamics[16-19]. Integral and off-policy reinforcement learning have further been introduced into electrical-machine control[20-22]. More recently, an integral-reinforcement-learning-based robust control scheme has demonstrated data-driven \(H_\infty\) regulation of PMSMs without requiring prior motor-parameter knowledge[23]. These studies show that robust feedback policies can be obtained directly from operating data rather than from explicitly identified motor models. However, when a data-driven feedback controller is applied directly to the original PMSM control problem, the learning process must address the nominal plant dynamics, parameter variations, coupling effects, and disturbances simultaneously. Meanwhile, disturbance-observer-based \(H_\infty\) control has also been investigated for PMSM drives[2], showing the effectiveness of combining disturbance compensation with robust feedback regulation, although the \(H_\infty\) controller in such approaches remains model-based.

These observations motivate a different control philosophy: rather than learning the entire PMSM current-control problem directly, the dominant unknown dynamics can first be compensated through a model-free feedforward channel, leaving only the post-compensation residual dynamics to the data-driven robust feedback controller. Based on this idea, this paper proposes a model-free PMSM current-control method based on a functional decomposition into disturbance feedforward and residual feedback. An ultra-local current model incorporates motor dynamics, parameter variations, cross-coupling effects, and other nonideal factors into generalized lumped disturbances. An extended state observer (ESO) is then employed to estimate these lumped disturbances and generate the feedforward compensation, thereby transforming the original PMSM current-control problem into the regulation of a simplified post-compensation residual system. Rather than designing the residual controller from a nominal motor model or using a manually selected fixed feedback gain, a state-feedback \(H_\infty\) policy is obtained directly from operating data through off-policy integral reinforcement learning. Since the learning process is performed on the simplified residual system, the value function and control policies can be parameterized in quadratic and linear forms, respectively, reducing the off-policy learning problem to low-dimensional parameter estimation without requiring neural-network approximation. In addition, electrical-parameter variations and input-gain mismatch are incorporated into the ESO-estimated lumped dynamics, whereas the residual-feedback policy is obtained from operating data. Consequently, neither prior knowledge nor online identification of PMSM electrical parameters is required for controller synthesis.

The main contributions of this work are summarized as follows:
\begin{itemize}[label=\raisebox{0.2ex}{\scalebox{0.6}{$\blacksquare$}},leftmargin=*,itemsep=0pt,parsep=0pt]
\item A model-free PMSM current-control method based on disturbance-feedforward and residual-feedback decomposition is developed. Unlike methods that directly regulate the original PMSM current dynamics with a single model-based or data-driven controller, the proposed method first compensates the dominant lumped dynamics through ESO-based feedforward action and subsequently formulates the post-compensation dynamics as a separate residual-feedback problem.

\item Motor-parameter-independent controller synthesis is achieved without explicit electrical-parameter identification. Electrical-parameter variations and input-gain mismatch are incorporated into the generalized lumped dynamics estimated by the ESO, while the residual \(H_\infty\) controller is obtained directly from operating data. Hence, neither prior knowledge nor online identification of stator resistance, stator inductance, or the permanent-magnet flux linkage is required.

\item A low-dimensional data-driven \(H_\infty\) solution is established for the post-compensation residual system. The simplified residual dynamics permit quadratic value-function and linear-policy parameterizations, avoiding neural-network approximation. Comparative simulations further show that, once learned, the feedback policy can be kept fixed without retraining or motor-specific controller retuning while maintaining nearly unchanged current-control performance under large variations in PMSM electrical parameters.
\end{itemize}

The remainder of this paper is organized as follows. Section 2 establishes the ultra-local current-loop formulation and introduces the disturbance-feedforward and residual-feedback decomposition. Section 3 develops the ESO-based disturbance-feedforward controller and analyzes the influence of the input scaling factor. Section 4 formulates the post-compensation residual system as a data-driven \(H_\infty\) control problem and derives the corresponding off-policy learning solution. Section 5 presents the simulation results and comparative performance evaluation. Finally, Section 6 concludes the paper.

\section{Model-Free Current-Loop Formulation and Feedforward-Residual-Feedback Decomposition}
\label{sec:headings}
The mathematical model serves as the prerequisite for theoretical analysis and simulation modeling, and constitutes the most fundamental step prior to the research of control strategies. Traditional modeling methods aim to develop high-precision and error-free models, which explore both known and unknown system dynamics through physical analysis, frequency sweeping and other approaches. However, the models established in this way tend to be complex and overly idealized, making them unsuitable for direct control strategy design. As a widely studied topic in recent years, model-free control has introduced the concept of the ultra-local model. This type of model only takes input and output signals into account while ignoring the intricate internal dynamics of the system, which greatly reduces model complexity and lays a solid foundation for the design of concise and efficient controllers.
\subsection{Ultra-Local Model-Free Control}
For a single-input single-output first-order nonlinear system, an ultra-local model can be established using only the input and output signals.
\begin{equation}
\dot{y} = F + \alpha u
\end{equation}
where \(y\) and \(u\) represent the output and input of the system, respectively. The parameter \(\alpha \) is an artificially introduced numerical scaling factor used to match the magnitudes of the input and output, while \(F\) is the unknown system dynamics, including unmodeled dynamics and external disturbances.
Based on (1), a generic model-free tracking control law can be expressed as
\begin{equation}
u = \frac{-\hat{F} + \dot{y}^{ref} + f(e)}{\alpha}
\end{equation}
where \(\hat{F}\) is the estimation of \(F\), \(y^{ref}\) is the reference output, and \(e = y - y^{ref}\) is the tracking error, \(f(e)\) represents an arbitrary function of current and past error variables. Under the ideal estimation condition  \(\hat{F} = F\),  substituting (2) into (1) gives
\begin{equation}
\dot{e} - f(e) = 0
\end{equation}
Choosing a proportional feedback function, \(f(e)=-K_p e (Kp>0)\), yields a standard first-order linear system

\begin{equation}
\dot{e} + K_p e = 0
\end{equation}
The analytical solution of this first-order linear homogeneous differential equation is
\begin{equation}
e(t) = e(0) \cdot e^{-K_p t}
\end{equation}
Therefore, as long as the proportional coefficient \(K_p\) is greater than 0, the tracking error will converge to zero exponentially regardless of the initial error \(e(0)\). The convergence rate is entirely determined by \(K_p\): a larger \(K_p\) leads to faster error convergence and a higher system response bandwidth. Ultimately, by means of feedforward compensation for \(\hat{F}\), the model-free control simplifies the closed-loop error dynamics into a first-order linear system completely independent of the original plant. Zero steady-state tracking error can be achieved merely with a proportional element.
\subsection{Ultra-Local Model of the PMSM Current Loop}
The stator voltage equation in the d-q frame is expressed as
\begin{equation}
\begin{cases}
u_d = R_s i_d + \dfrac{d\varphi_d}{dt} - \omega_e \varphi_q \\[8pt]
u_q = R_s i_q + \dfrac{d\varphi_q}{dt} + \omega_e \varphi_d
\end{cases}
\end{equation}
In the d-q coordinate system, the stator flux linkage equations read
\begin{equation}
\begin{cases}
\varphi_d = L_d i_d + \varphi_f \\[4pt]
\varphi_q = L_q i_q
\end{cases}
\end{equation}
 By substituting (7) into (6)
\begin{equation}
\begin{cases}
u_d = R_s i_d + L_d \dfrac{di_d}{dt} - \omega_e L_q i_q \\[8pt]
u_q = R_s i_q + L_q \dfrac{di_q}{dt} + \omega_e L_d i_d + \omega_e \varphi_f    
\end{cases}
\end{equation}
The state-space equation of the fundamental current loop is obtained from the voltage equation (8)
\begin{equation}
\begin{cases}
\dfrac{di_d}{dt} = \dfrac{1}{L_d} u_d -\dfrac{R_s}{L_d} i_d + \dfrac{L_q}{L_d} \omega_e i_q -\dfrac{1}{L_d} \Delta u_d \\[8pt]
\dfrac{di_q}{dt} = \dfrac{1}{L_q} u_q -\dfrac{R_s}{L_q} i_q - \dfrac{1}{L_q} \omega_e (L_d i_d + \varphi_f) - \dfrac{1}{L_q} \Delta u_q    
\end{cases}
\end{equation}
where \(\Delta u_d\) and \(\Delta u_q\) represent the voltage disturbances caused by the uncertainties of the system parameters and the inverter and nonlinearities. Traditional PI controllers neglect the cross-coupling terms and simplify the equation into a second-order linear system for control law design. Furthermore, the selection of proportional and integral gains relies on the resistance and inductance parameters of the model, which brings many drawbacks in practical applications.
Therefore, it is necessary to establish a new model for the controlled plant. The ultra-local model of the current loop derived from model-free control theory is given
\begin{equation}
\begin{cases}
\dot{i}_d= F_d + \alpha_{sd} u_d \\[8pt]
\dot{i}_q = F_q + \alpha_{sq} u_q
\end{cases}
\end{equation}
where the parameter \(\alpha_s\) is the scaling factor of the input, and \(F_d\) and \(F_q\) represent the lumped disturbances in the d-q axes, including known terms related to motor parameters, unknown voltage disturbances and other non-ideal factors. Based on the ultra-local current model in (10), the control law is formulated as
\begin{equation}
\begin{cases}
u_d = \dfrac{-\hat{F}_d + \dot{i_d}^{ref} + f(e_d)}{\alpha_{sd}} \\[8pt]
u_q = \dfrac{-\hat{F}_q + \dot{i_q}^{ref} + f(e_q)}{\alpha_{sq}}
\end{cases}
\end{equation} 

where \(\hat F_d\) and \(\hat F_q\) are the estimates of the lumped dynamics \(F_d\) and \(F_q\), respectively, while \(i_d^{ref}\) and \(i_q^{ref}\) denote the d- and q-axis current references. It follows (11) that the control input admits a natural functional decomposition into feedforward and residual-feedback components. Specifically, the estimated lumped dynamics are used to construct the disturbance-compensation feedforward action, whereas the derivative of the prescribed current reference constitutes a reference-derivative feedforward action. The remaining error-dependent control action is assigned to the residual-feedback channel. Accordingly, the control input can be expressed as
\begin{equation}
\begin{cases}
u_d = u_d^{ff} + u_d^{fb} = \underbrace{-\frac{\hat{F}_d}{\alpha_{sd}}}_{u_{d,\mathrm{dis}}^{ff}}+ \underbrace{\frac{\dot{i}_d^{\mathrm{ref}}}{\alpha_{sd}}}_{u_{d,\mathrm{ref}}^{ff}}+ u_d^{fb}  \\[9pt]
u_q = u_q^{ff} + u_q^{fb} = \underbrace{-\frac{\hat{F}_q}{\alpha_{sq}}}_{u_{q,\mathrm{dis}}^{ff}}+ \underbrace{\frac{\dot{i}_q^{\mathrm{ref}}}{\alpha_{sq}}}_{u_{q,\mathrm{ref}}^{ff}}+ u_q^{fb}
\end{cases}
\end{equation}  

where \(u_{d,\mathrm{dis}}^{ff}\) and \(u_{q,\mathrm{dis}}^{ff}\) denote the disturbance-compensation feedforward components, \(u_{d,\mathrm{ref}}^{ff}\) and \(u_{q,\mathrm{ref}}^{ff}\) denote the reference-derivative feedforward components,  \(u_d^{ff}\) and \(u_q^{ff}\) denote the feedforward components, and \(u_d^{fb}\) and \(u_q^{fb}\) denote the residual-feedback components.

Substituting the feedforward control law into (10)  yields
\begin{equation}
\begin{cases}
\dot{e}_d = \Delta F_d + {\alpha_{sd}}u_d^{fb}\\[8pt]
\dot{e}_q = \Delta F_q + {\alpha_{sq}}u_q^{fb}
\end{cases}
\end{equation}

where the disturbance residual \(\Delta F_d = F_d - \hat{F}_d\), \(\Delta F_q = F_q - \hat{F}_q\). The tracking error \(e_d = i_d - i_d^{ref} \), \(e_q = i_q - i_q^{ref}\).

A proper feedback control law designed for the system to eliminate residuals and combined with feedforward compensation for lumped disturbances, accurate current tracking can be achieved.
\section{ESO-Based Disturbance Feedforward Compensation}
In the ultra-local current model, feedforward compensation of the lumped dynamics provides the basis for the subsequent regulation of the post-compensation residual system. Therefore, accurate estimation of the lumped dynamics is essential to the proposed control method. Various estimation approaches can be employed, including algebraic identification, recursive least squares, and observer-based methods. Among them, the extended state observer (ESO) is adopted in this work owing to its simple structure and strong capability for estimating generalized disturbances and unmodeled dynamics. Since the d- and q-axis ultra-local current models share the same mathematical structure, the following ESO design and feedforward analysis are developed for a generic current axis. The axis subscript is omitted for compactness, and all results apply identically to both axes.
\subsection{ESO Design and Discrete-Time Stability}
Based on the ultra-local current model in (10), a linear extended state observer is constructed with the current and lumped dynamics as extended states. The observer is formulated as
\begin{equation}
\begin{cases}
\dot{\hat{i}} = \hat{F} + \alpha_s u + h_1 (i -\hat{i}) \\[8pt]
\dot{\hat{F}} = h_2 (i - \hat{i})
\end{cases}
\end{equation}
where \(\hat{i}\) and \(\hat{F}\) are the estimations value of \(i\) and \(F\), \(h_1\) and \(h_2\) denote the output feedback coefficients for the observer. The system block diagram of the disturbance observer can be constructed according to (14), as shown in Figure 1.
\begin{figure}[htbp]
  \centering
  \includegraphics[width = 0.58\textwidth]{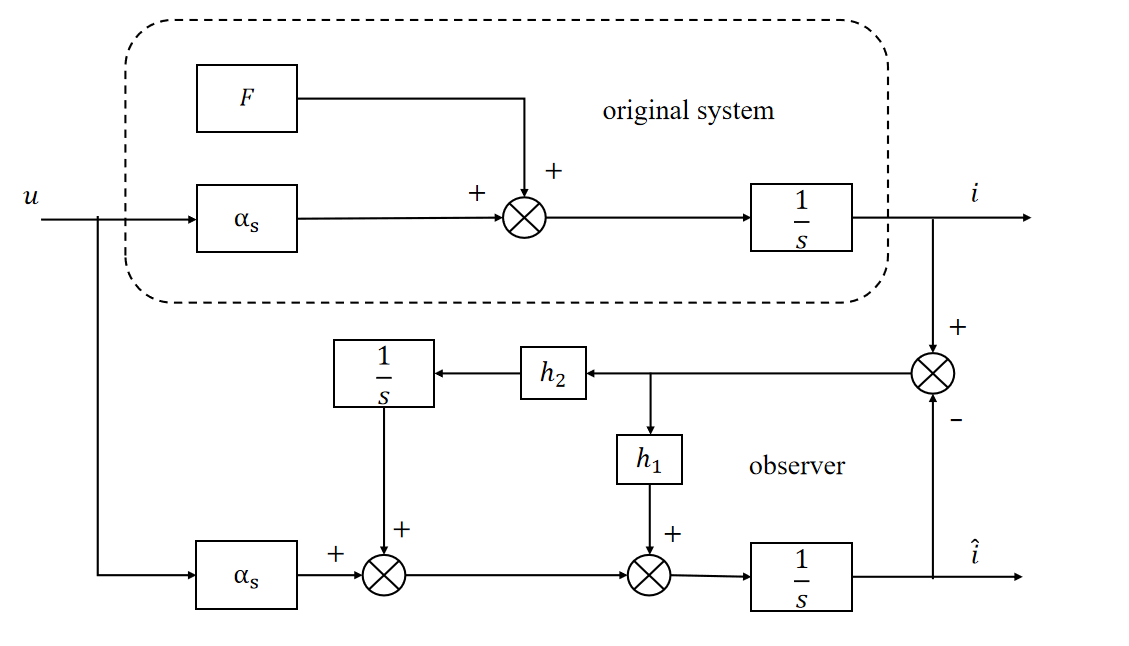}
  \caption{Block diagram of the ultra-local ESO for a generic current axis}
  \label{fig:fig1}
\end{figure}

When the observed current converges to its true value, the observed disturbance will also converge to the true value. The selection of \(h_1\) and \(h_2\) determines the stability and convergence rate of the observer. To design these two parameters, (14) is rewritten in matrix form
\begin{equation}
\begin{cases}
  \dot{\hat{x}} = \mathbf{A} \hat{x} + \mathbf{B} u + \mathbf{H} (y-\hat{y})\\
  \hat{y} = \mathbf{C} \hat{x}
\end{cases}
\end{equation}
where \(\hat{x} = [\hat{i}, \hat{F}]^T\), \(u = u\), \(y = i\), and \(\mathbf{A} = \begin{bmatrix} 0 & 1 \\[8pt] 0 & 0 \end{bmatrix}\), \(\mathbf{B} = \begin{bmatrix} \alpha_s \\[8pt] 0 \end{bmatrix}\), \(\mathbf{C} = \begin{bmatrix} 1 & 0 \end{bmatrix}\) and the observer gain matrix is \(\mathbf{H} = \begin{bmatrix} h_1 \\[8pt] h_2 \end{bmatrix}\). The characteristic equation of the system is given by
\begin{equation}
  \det(s \mathbf{I} - (\mathbf{A} - \mathbf{H} \mathbf{C})) = s^2 + h_1 s + h_2 = 0
\end{equation}
To facilitate analysis, repeated roots \(s_1 = s_2 = -\omega_0\) are designed for the system. The formulas of gain coefficients are derived by inverse deduction
\begin{equation}
\begin{cases}
  h_1 = 2 \omega_0 \\[8pt]
  h_2 = \omega_0^2
\end{cases}
\end{equation}
A continuously stable observer is achievable with a positive \(\omega_0\). This parameter represents the observer bandwidth, and a larger \(\omega_0\) leads to a faster response. However, this is an analysis under ideal assumptions. When the observer is applied to discrete sampled digital control systems, the selection of \(\omega_0\) is also constrained by the sampling period \(T_\mathrm{s}\). Discretizing (15) via first-order Taylor expansion yields
\begin{equation}
  \hat{x}(k+1) = \mathbf{A}_z \hat{x}(k) + \mathbf{B}_z u(k) + T_\mathrm{s} \mathbf{H} y(k) 
\end{equation}
where \(\mathbf{A}_z =  T_\mathrm{s} \mathbf{A} + \mathbf{I} - T_\mathrm{s} \mathbf{H} \mathbf{C}\) is discrete system matrix and \(\mathbf{B}_z = T_\mathrm{s} \mathbf{B}\) is the discrete input matrix. The characteristic equation of the discrete system is given by
\begin{equation}
\det(z \mathbf{I} - \mathbf{A}_z) = \det(z \mathbf{I} - (T_\mathrm{s} \mathbf{A} + \mathbf{I} - T_\mathrm{s} \mathbf{H} \mathbf{C})) = 0
\end{equation}
thus
\begin{equation}
  z^2 + (h_1 T_\mathrm{s}-2) z - h_1 T_\mathrm{s} + h_2 T_\mathrm{s}^2 + 1 = 0
\end{equation}
After substitution into (17), the equation can be simplified to
\begin{equation}
  (z + (\omega_0 T_{s} - 1))^2 = 0
\end{equation}

Under the forward-Euler discretization, the observer poles are given by \(z_1 = z_2 = 1-\omega_0 T_{s}\). Therefore, asymptotic stability requires
\begin{equation}
  0 < \omega_0 T_{s} <  2
\end{equation}
The observer bandwidth \(\omega_0\) determines the convergence rate of the estimation error. Moving the poles toward the origin generally accelerates the nominal observer response, whereas an excessively large bandwidth may increase sensitivity to measurement noise, discretization errors, and unmodeled high-frequency dynamics. Therefore, \(\omega_0\)  should be selected within the stability range by balancing estimation speed and practical robustness.
\subsection{Feedforward Compensation and Scaling-Factor Mismatch Analysis}
The scaling factor \(\alpha_s\) in the ultra-local model is an artificially selected numerical coefficient and is not assumed a priori to coincide with the physical current-input gain of the PMSM. To clarify the influence of scaling-factor mismatch, let \(\alpha_0\)  denote the actual current-input gain, which equals the reciprocal of the stator inductance for the ideal PMSM current model, and let \(u_\Sigma\) denote the total voltage applied to the plant. For either current axis, the actual plant and the ESO can be expressed as
\begin{equation}
\begin{aligned}
\dot{i} &= F_0 + \alpha_0 u_\Sigma\\[8pt]
\dot{\hat{i}} &= \hat{F} + \alpha_s u_\Sigma + h_1(i - \hat{i})
\end{aligned}
\end{equation}
where \(F_0\) denotes the lumped dynamics defined with respect to \(\alpha_0\). Defining the ESO current-estimation error as \(\Delta i = i - \hat{i}\), the two equations yield
\begin{equation}
\hat{F} = F_0+(\alpha_0-\alpha_s) u_\Sigma-h_1 \Delta i-\Delta\dot{i}
\end{equation}
As shown in (24), the input-gain mismatch is naturally incorporated into the ESO-estimated lumped dynamics. Once the current estimate converges, i.e., \(\Delta i \rightarrow 0,\Delta \dot{i} \rightarrow 0  \), using \(F_0 = \dot{i} - \alpha_0 u_\Sigma \)gives the fundamental relation
\begin{equation}
  \hat{F} = \dot{i} - \alpha_s u_\Sigma
\end{equation}
Using (25), the influence of the selected scaling factor on the two feedforward formulations can be compactly expressed as
\begin{equation}
\begin{cases}
\Delta_\alpha u_{\mathrm{dis}}^{ff} = -\dot{i}\left(\dfrac{1}{\alpha_s}-\dfrac{1}{\alpha_0}\right), & u_{\mathrm{dis}}^{ff}=-\dfrac{\hat{F}}{\alpha_s},\\[6pt]
\Delta_\alpha u^{ff} = -\dot{e}\left(\dfrac{1}{\alpha_s}-\dfrac{1}{\alpha_0}\right), & u^{ff}=\dfrac{-\hat{F}+\dot{i}^{\mathrm{ref}}}{\alpha_s}.
\end{cases}
\end{equation}
where \(\Delta_\alpha u_{\mathrm{dis}}^{ff} \triangleq u_\mathrm{{dis}}^{ff} (\alpha_s )-u_\mathrm{{dis}}^{ff} (\alpha_0 )\), and the comparison is made for the same physical operating trajectory. (26) shows that the scaling-factor-induced discrepancy of the disturbance-compensation component is proportional to the current derivative (\(\dot{i}\)), whereas that of the complete feedforward action is proportional to the tracking-error derivative (\(\dot{e}\)). Consequently,
\begin{equation}
\dot{i}=0 \Rightarrow \Delta_\alpha u_{\mathrm{dis}}^{ff}=0,\quad \dot{e}=0 \Rightarrow \Delta_\alpha u^{ff}=0.
\end{equation}
As demonstrated by (26) and (27), the explicit appearance of \(\alpha_s\) in the feedforward laws does not imply dependence on an accurately identified physical input gain. For the disturbance-compensation component, scaling-factor invariance is obtained at current steady state (\(\dot i=0\)); with the reference-derivative component included, the corresponding condition is relaxed to \(\dot e=0\). Therefore, when the tracking-error dynamics converge, the complete feedforward action becomes independent of the selected \(\alpha_s\), and \(\alpha_s\) is not required to accurately match the physical gain \(\alpha_0=1/L_s\). Thus, accurate knowledge of the stator inductance is not required for feedforward synthesis. During transients, finite ESO dynamics, tracking-error dynamics, sampling effects, and other nonidealities may still produce scaling-factor-dependent discrepancies, which appear as post-compensation residual dynamics and motivate the residual-feedback controller developed in the following section.

\section{Data-Driven \(H_\infty \) Residual Feedback Control}
The preceding section develops the ESO-based feedforward compensation for the generalized lumped dynamics. However, feedforward action alone does not provide closed-loop stabilization or robustness against post-compensation residuals. In practical operation, finite observer bandwidth, delay, saturation, and other nonideal effects may result in imperfect disturbance compensation. Therefore, a feedback channel is required to regulate the remaining residual dynamics and suppress tracking errors. In this work, this task is addressed by a data-driven \(H_\infty\) residual-feedback controller.

In Section 2.2, (13) characterizes the residual system compensated by disturbance feedforward, and this section aims to design a feedback control law for the system. Fliess suggested tuning the feedback term's order and parameters through experiments and trial and error method in model-free control[3]. Reference [7] introduces proportional error feedback in ultra-local MFPCC and demonstrates zero steady-state current-tracking error under parameter mismatch, while Reference [12] shows that the controller gain may still depend on motor inductance. This motivates the proposed data-driven \(H_\infty\) residual feedback, which avoids motor-specific gain tuning and explicit parameter dependence.
\subsection{\(H_\infty\) Formulation of the Post-Compensation Residual System}
As a robust control approach, state-feedback \(H_\infty \) control can suppress disturbances while balancing tracking errors and control efforts, which naturally matches the disturbance residual model proposed above. The model is reformulated as follows
\begin{equation}
  \dot{x} = F(x) + G u + K w
\end{equation}
where state variable \(x = [e_d, e_q]^T \), input \(u = [u_d^{fb}, u_q^{fb}]^T\), and disturbance \(w = [\Delta F_d, \Delta F_q]^T\). The internal system dynamics satisfy \(F(\boldsymbol{x}) = \boldsymbol{0}\), the input matrix is \(\boldsymbol{G} = diag(\alpha_{sd},\alpha_{sq}) \), and the disturbance matrix is \(\boldsymbol{K} = \boldsymbol{I}_2\). The state-feedback \(H_\infty\) problem aims to seek a state feedback control law such that the closed-loop system is asymptotically stable with its \(L_2\) gain no greater than \(\gamma\)[24]. This requirement can be described by the following equations
\begin{equation}
\begin{gathered}
\int_{0}^{\infty} \left(x^T Q x + u^T R u\right) \mathrm{d}\tau \le \gamma^2 \int_{0}^{\infty} w^T w \mathrm{d}\tau \\
x(0) = 0,\quad w\in L_2[0,\infty).
\end{gathered}
\label{eq:hinf_performance}
\end{equation}

where \(Q\) and \(R\) are positive definite matrices, and \(\gamma\) is the upper bound of the \(L_2\) gain and reflects the system's ability to attenuate bounded disturbances. The state-feedback \(H_\infty\) control law can be designed to satisfy this performance criterion. This is essentially a two-player zero-sum game problem. The objective is to derive the state feedback control law \(\boldsymbol{u}(\boldsymbol{x})\) to minimize tracking errors and control effort consumption for optimal system performance. By contrast, the adversary intends to maximize the tracking error via the disturbance law \(\boldsymbol{w}(\boldsymbol{x})\), deteriorate system stability and trigger the worst-case performance degradation. An optimal solution subject to prescribed robustness constraints is finally obtained from this game interaction. The cost function is formulated as
\begin{equation}
  J(\boldsymbol{u}, \boldsymbol{w}) =\int_0^\infty (x^T Q x + u^T R u - \gamma^2 w^T w) d\tau
\end{equation} 
According to the dynamic-programming principle, \(V^*(x)\) satisfies the Hamilton-Jacobi-Isaacs (HJI) equation as
\begin{equation}
  H(V^{*}, u, w) = x^T Q x + u^T R u - \gamma^2 w^T w + (V_x^*)^T (F(x) + G u + K w) = 0
\end{equation}

According to the extremum conditions, set \(\partial H/\partial \boldsymbol{u} = \boldsymbol{0}\) and \(\partial H/\partial \boldsymbol{w} = \boldsymbol{0}\) respectively, and the optimal policies are derived as follows
\begin{equation}
  \begin{cases}
    u^* = -\frac{1}{2} R^{-1} G^T V_x^* \\[8pt]
    w^* = \frac{1}{2 \gamma^2} K^T V_x^* 
  \end{cases}
\end{equation}  
Finally, substitute the optimal control law \(\boldsymbol{u}^{*}\) and the worst-case disturbance law \(\boldsymbol{w}^{*}\) into (31), eliminate \(\boldsymbol{u}\) and \(\boldsymbol{w}\), yield the Hamilton-Jacobi-Isaacs (HJI) equation
\begin{equation}
  H(V^{*}, u^*, w^*) = x^T Q x + (V_x^*)^T F(x) - \frac{1}{4} (V_x^*)^T G R^{-1} G^T V_x^* + \frac{1}{4 \gamma^2} (V_x^*)^T K K^T V_x^* = 0
\end{equation}  
For nonlinear systems, the HJI equation is a nonlinear partial differential equation that admits no analytical solutions in general. So it is commonly solved via data-driven approaches.
However, for linear time-invariant systems \(\dot{\boldsymbol{x}} = A\boldsymbol{x} + G\boldsymbol{u} + K\boldsymbol{w}\), the optimal value function takes a quadratic form \(V^{*}(\boldsymbol{x}) = \boldsymbol{x}^{T}P\boldsymbol{x}\). The analytically solvable Algebraic Riccati Equation (ARE) can be obtained by degenerating the HJI equation
\begin{equation}
  A^T P + P A  + Q - P G R^{-1} G^T P + \frac{1}{\gamma^2} P K K^T P = 0
\end{equation}
Once P is solved, the optimal control law reads \(\boldsymbol{u}^{*} = -R^{-1}G^{T}P\boldsymbol{x}\).
\subsection{Off-Policy Integral Reinforcement Learning for Data-Driven \(H_\infty \) Control}
Although the nominal post-compensation residual system has a simple structure, direct solution of the \(H_\infty\) algebraic Riccati equation requires explicit knowledge of the residual-system dynamics and the input and disturbance matrices. In particular, the input matrix \(G\) contains the ultra-local scaling factors, while practical nonidealities may introduce additional uncertainty into the residual dynamics. Explicit identification of these quantities would reintroduce model dependence into the controller design. Therefore, an off-policy integral reinforcement learning (IRL) method [18] is adopted to obtain the \(H_\infty\) feedback policy directly from operating data without explicitly identifying the residual-system matrices. The corresponding off-policy learning structure is illustrated in Figure 2.
\begin{figure}[htbp]
  \centering
  \includegraphics[width = 0.62\textwidth]{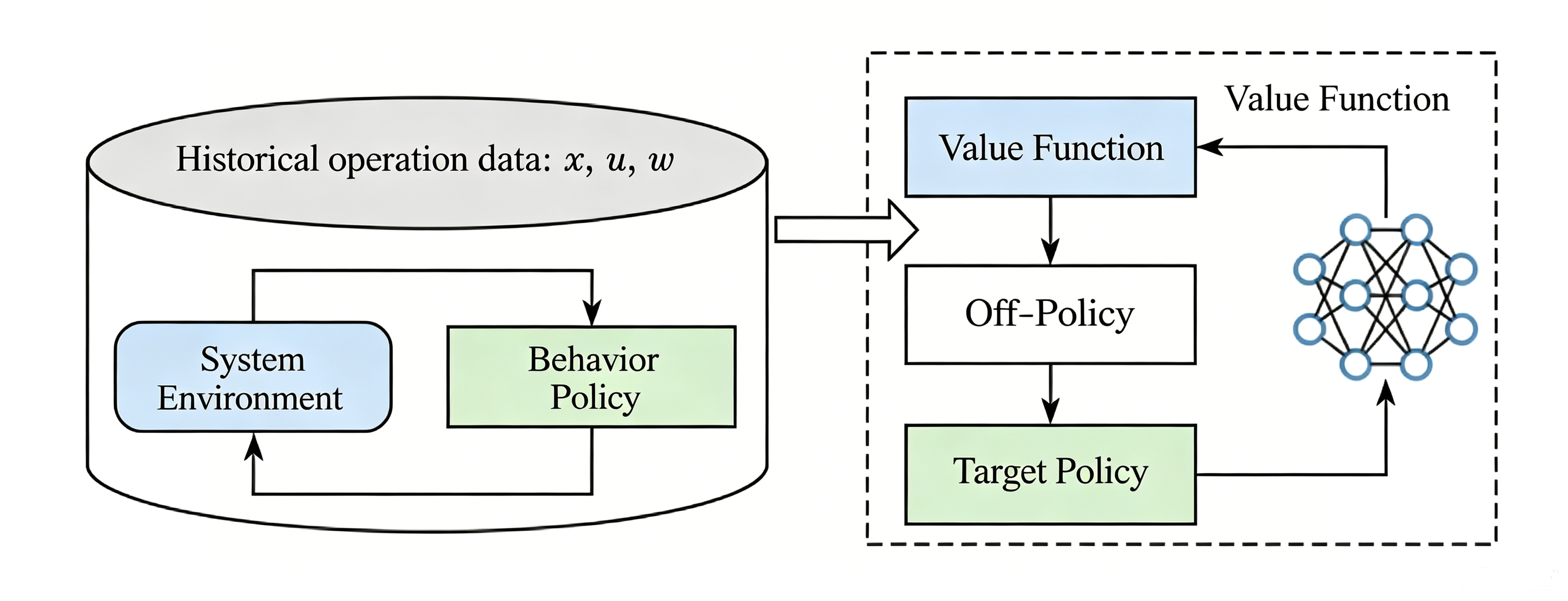}
  \caption{Off-policy learning framework for data-driven \(H_\infty\) residual feedback}
  \label{fig:fig2}
\end{figure}

Rewriting the residual system in (28) around the target policies \(u^i\) and \(w^i\) gives
\begin{equation}
  \dot{x} = F(x) + G u^{i} + K w^{i} + G (u - u^{i}) + K (w - w^{i})
\end{equation}
where \(u^{i}\) and \(w^{i}\) are the target policy at the i-th iteration, which is iteratively trained and optimized. \(\boldsymbol{u}\) and \(\boldsymbol{w}\) are the behavior policy, which serves as the initial policy for data collection. The Hamiltonian function at the i-th iteration is expressed as
\begin{equation}
  x^T Q x + (u^{i})^T R u^{i} - \gamma^2 (w^{i})^T w^{i} + (V_x^{i})^T (F(x) + G u^{i} + K w^{i})= 0
\end{equation}
Multiply both sides of (35) by \(\left(V_{x}^{i}\right)^{T}\), then substitute (36) into the resulting expression to eliminate the term \(F(x)\)
\begin{equation}
  \dot{V^{i}} = -x^T Q x - (u^{i})^T R u^{i} + \gamma^2 (w^{i})^T w^{i} + (V_x^{i})^T G (u - u^{i}) + (V_x^{i})^T K (w - w^{i})
\end{equation}
Next, the model-dependent matrices G and K will be eliminated through algebraic transformation. First, integrate both sides of (37) over the interval \([t, T+t]\)
\begin{equation}
  V^{i}(x(T+t)) - V^{i}(x(t)) = \int_{t}^{t+T} (\gamma^2 (w^{i})^T w^{i} - x^T Q x - (u^{i})^T R u^{i}) dt + \int_{t}^{t+T} (V_x^{i})^T (G (u - u^{i}) + K (w - w^{i})) dt
\end{equation}
 The policies \(\boldsymbol{u}^{i+1}\) and \(\boldsymbol{w}^{i+1}\) required for the next iteration are theoretically computed from the current value function \(V^{i}\). Referring to (32), it can be known that
\begin{equation}
  \begin{cases}
    u^{i+1} = -\frac{1}{2} R^{-1} G^T V_x^{i} \\[8pt]
    w^{i+1} = \frac{1}{2 \gamma^2} K^T V_x^{i} 
  \end{cases}
\end{equation}
Rearranging this expression yields
\begin{equation}
    \begin{cases}
    (V_x^{i})^T G = -2 (u^{i+1})^T R\\[8pt]
    (V_x^{i})^T K = 2 (w^{i+1})^T \gamma^2 
  \end{cases}
\end{equation}
Subsituting (40) into (38) yields
\begin{equation}
  \begin{aligned}
    V^{i}(x(t+T)) - V^{i}(x(t))
    &= \int_{t}^{t+T} \left( \gamma^{2}(w^{i})^{T}w^{i} - x^{T}Qx - (u^{i})^{T}Ru^{i} \right) dt \\
    &\quad - \int_{t}^{t+T} 2(u^{i+1})^{T}R\left(u - u^{i}\right) dt \\
    &\quad + \int_{t}^{t+T} 2(w^{i+1})^{T}\gamma^{2}\left(w - w^{i}\right) dt
  \end{aligned}
\end{equation}

\subsection{Low-Dimensional Policy Parameterization and Data-Driven Solution}
At this point, all model-related information contained in the HJI equation has been completely eliminated. Note that \(\boldsymbol{u}^{i+1}\) and \(\boldsymbol{w}^{i+1}\) represent future policies, which remain unknown during the current iteration. Additionally, the form of the value function \(V^{i}(\boldsymbol{x})\) is also unavailable. For this reason, a neural network structure is generally adopted to approximate the value function, as illustrated in Figure 3.
\begin{figure}[htbp]
  \centering
  \includegraphics[width = 0.62\textwidth]{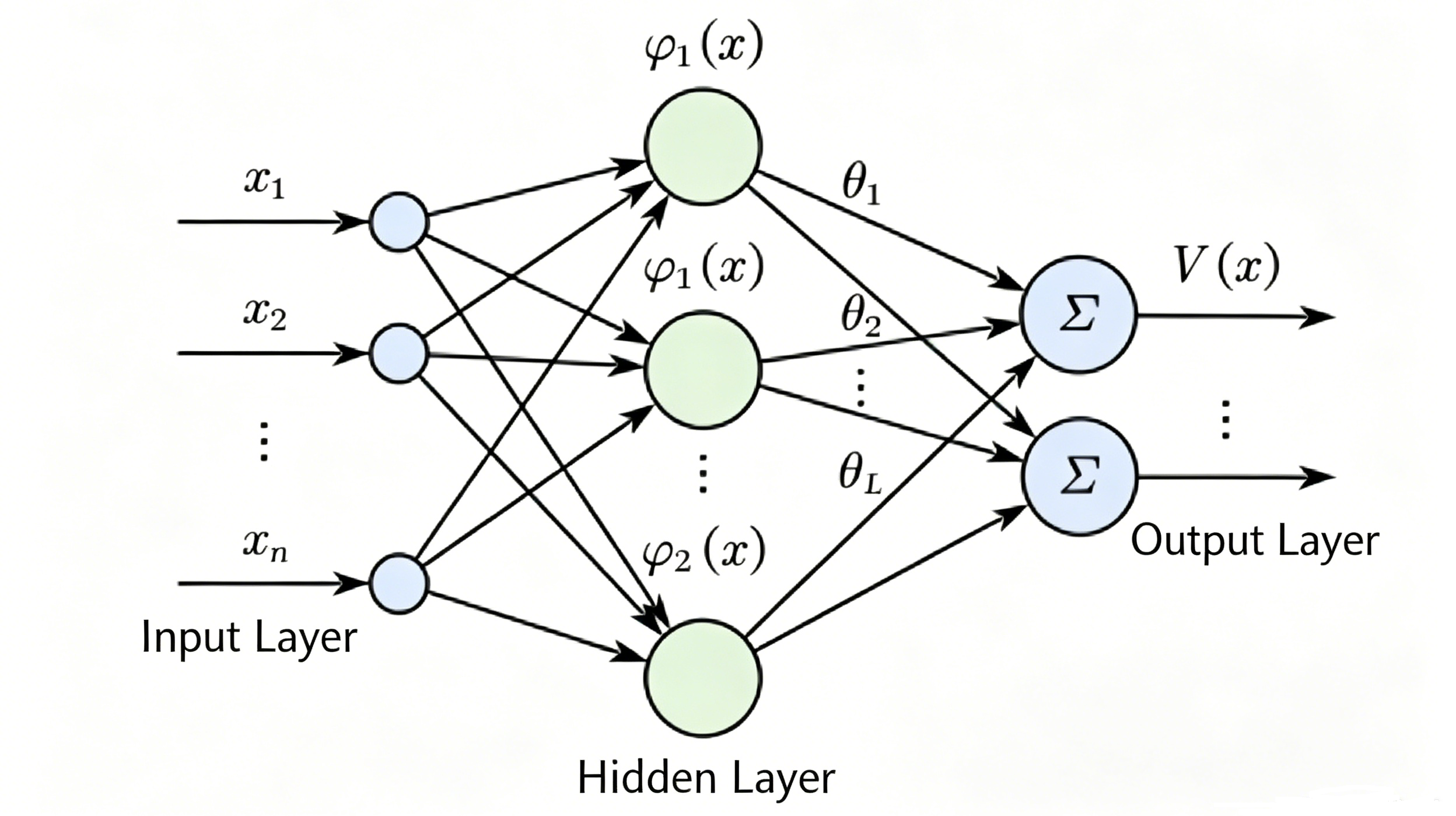}
  \caption{Generic neural-network structure for value-function approximation}
  \label{fig:fig3}
\end{figure}

Nevertheless, training neural networks imposes a substantial overhead on the microprocessors used for motor control, which demands abundant storage space and sufficient on-chip computing power.
Because of the linear characteristic of System (28), this paper assumes that \(V^{i}(\boldsymbol{x})\) is a quadratic form of the state variable \(\boldsymbol{x}\), and the policies \(\boldsymbol{u}^{i+1}\) and \(\boldsymbol{w}^{i+1}\) are approximated as state feedback. Taking the two-dimensional state vector \(\boldsymbol{x} = \begin{bmatrix} x_1 & x_2 \end{bmatrix}^{T}\) as an example, the value function can be expressed as
\begin{equation}
  V^{i}(x) = x^T \boldsymbol{\theta }_v x = \begin{bmatrix} x_1 & x_2 \end{bmatrix} \begin{bmatrix} \theta_{v}^{11} & \theta_{v}^{12} \\[8pt] \theta_{v}^{21} & \theta_{v}^{22} \end{bmatrix} \begin{bmatrix} x_1 \\[8pt] x_2 \end{bmatrix}
\end{equation}
The control law and disturbance law can be expressed as
\begin{equation}
    u^{i+1} = \boldsymbol{\theta }_u x =\begin{bmatrix} \theta_{u_d}^{1} & \theta_{u_d}^{2} \\[8pt] \theta_{u_q}^{1} & \theta_{u_q}^{2} \end{bmatrix} \begin{bmatrix} x_1 \\[8pt] x_2 \end{bmatrix}
\end{equation}
\begin{equation}
    w^{i+1} = \boldsymbol{\theta }_w x =\begin{bmatrix} \theta_{w_d}^{1} & \theta_{w_d}^{2} \\[8pt] \theta_{w_q}^{1} & \theta_{w_q}^{2} \end{bmatrix} \begin{bmatrix} x_1 \\[8pt] x_2 \end{bmatrix} 
\end{equation}
Let the control weight matrix \(\boldsymbol{R} = \mathrm{diag}(r_d, r_q)\). Decomposing (41) yields
\begin{equation}
  \begin{aligned}
    V^{i}(x(t+T)) - V^{i}(x(t))
    &= \int_{t}^{t+T} \left( \gamma^{2}(w^{i})^{T}w^{i} - x^{T}Qx - (u^{i})^{T}Ru^{i} \right) dt \\
    &\quad - \int_{t}^{t+T} 2\Big(u_{d}^{i+1}r_{d}(u_{d} - u_{d}^{i}) + u_{q}^{i+1}r_{q}(u_{q} - u_{q}^{i})\Big) dt \\
    &\quad + \int_{t}^{t+T} 2\gamma^{2}\Big(w_{d}^{i+1}(w_{d} - w_{d}^{i}) + w_{q}^{i+1}(w_{q} - w_{q}^{i})\Big) dt
  \end{aligned}
\end{equation}
 Rewrite \(\theta_v = \begin{bmatrix} \theta_{v}^{11} & 2\theta_{v}^{12}  & \theta_{v}^{22} \end{bmatrix}^{T}\), Rewrite \(\boldsymbol{V}^{i}(x)\), \(u^{i+1}(x)\) and \({w}^{i+1}({x})\) into vector inner product forms
\(\boldsymbol{V}^{i}(\boldsymbol{x}) = \boldsymbol{\theta}_{v}^{T}\boldsymbol{\varphi}(\boldsymbol{x}),\quad
\boldsymbol{u}_{d}^{i+1} = \boldsymbol{\theta}_{u_{d}}^{T}\boldsymbol{x},\quad
\boldsymbol{u}_{q}^{i+1}(\boldsymbol{x}) = \boldsymbol{\theta}_{u_{q}}^{T}\boldsymbol{x},\quad
\boldsymbol{w}_{d}^{i+1}(\boldsymbol{x}) = \boldsymbol{\theta}_{w_{d}}^{T}\boldsymbol{x},\quad
\boldsymbol{w}_{q}^{i+1}(\boldsymbol{x}) = \boldsymbol{\theta}_{w_{q}}^{T}\boldsymbol{x},\)
where \({\varphi}(x) = \begin{bmatrix} x_1^2 & x_1x_2 & x_2^2 \end{bmatrix}^{T}\). Substituting these into (45) yields
\begin{equation}
  \begin{aligned}
    \boldsymbol{\theta}_{v}^{T}\big(\boldsymbol{\varphi}(\boldsymbol{x}(t+T)) - \boldsymbol{\varphi}(\boldsymbol{x}(t))\big)
    &= \int_{t}^{t+T} \left( \gamma^{2}(\boldsymbol{w}^{i})^{T}\boldsymbol{w}^{i} - \boldsymbol{x}^{T}\boldsymbol{Q}\boldsymbol{x} - (\boldsymbol{u}^{i})^{T}\boldsymbol{R}\boldsymbol{u}^{i} \right) dt \\
    &\quad - \int_{t}^{t+T} 2\Big(\boldsymbol{\theta}_{u_d}^{T}\boldsymbol{x}r_d(u_d - u_d^{i}) + \boldsymbol{\theta}_{u_q}^{T}\boldsymbol{x}r_q(u_q - u_q^{i})\Big) dt \\
    &\quad + \int_{t}^{t+T} 2\gamma^{2}\Big(\boldsymbol{\theta}_{w_d}^{T}\boldsymbol{x}(w_d - w_d^{i}) + \boldsymbol{\theta}_{w_q}^{T}\boldsymbol{x}(w_q - w_q^{i})\Big) dt
  \end{aligned}
\end{equation}
At this point, all variables in the HJI equation are known and can be obtained in real time during iterations, except for the unknown parameter vector
\(\boldsymbol{\theta}_{L}^{i} = \begin{bmatrix}
\boldsymbol{\theta}_{v}^{T} & \boldsymbol{\theta}_{u_{d}}^{T} & \boldsymbol{\theta}_{u_{q}}^{T} & \boldsymbol{\theta}_{w_{d}}^{T} & \boldsymbol{\theta}_{w_{q}}^{T}
\end{bmatrix}^{T}\)
Therefore, \(\boldsymbol{\theta}_{L}^{i} \) can be solved via parameter estimation methods such as the recursive least squares(RLS) method and gradient descent method. For the convenience of derivation, batch least squares is adopted in this work. Rewrite the HJI equation into a matrix form as follows
\begin{equation}
  y(t)=(\boldsymbol{\theta}_{L}^{i})^T z(t) + \varepsilon (t)
\end{equation}
where
\begin{equation}
  y(t) = \int_{t}^{t+T} \left( \gamma^{2}(\boldsymbol{w}^{i})^{T}\boldsymbol{w}^{i} - \boldsymbol{x}^{T}\boldsymbol{Q}\boldsymbol{x} - (\boldsymbol{u}^{i})^{T}\boldsymbol{R}\boldsymbol{u}^{i} \right) dt 
\end{equation}
\begin{equation}
  z(t) = 
  \begin{bmatrix}
    \boldsymbol{\varphi}(\boldsymbol{x}(t+T)) - \boldsymbol{\varphi}(\boldsymbol{x}(t)) \\[8pt]
    \int_{\tau }^{\tau+T} 2r_d(u_d - u_d^{i}) x d\tau \\[8pt] 
    \int_{\tau}^{\tau+T} 2r_q(u_q - u_q^{i}) x d\tau \\[8pt]
    -\int_{\tau}^{\tau+T} 2\gamma^{2} (w_d - w_d^{i}) x d\tau \\[8pt]
    -\int_{\tau}^{\tau+T} 2\gamma^{2} (w_q - w_q^{i}) x d\tau
  \end{bmatrix}
\end{equation}
To guarantee sufficiently accurate parameter estimation, the error term \(\varepsilon(t)\) needs to be minimized as much as possible, which is also referred to as the TD residual in reinforcement learning.
The vector \(\boldsymbol{\theta}_{L}^{i}\) contains L weight parameters, so at least L groups of data samples need to be collected to satisfy the rank condition(\(N\geqslant L, rank(Z^i)=L=11\)). Suppose N groups of samples are collected in the i-th iteration with \(N \geq L\), the samples can be expressed as
\begin{equation}
 \left\{
\begin{aligned}
\boldsymbol{Y}^{i} &= \begin{bmatrix}
y(t+T) & y(t+2T) & y(t+3T) & \dots & y(t+NT)
\end{bmatrix}^{T} \\
\boldsymbol{Z}^{i} &= \begin{bmatrix}
z(t+T) & z(t+2T) & z(t+3T) & \dots & z(t+NT)
\end{bmatrix}^{T}
\end{aligned}
\right.
\end{equation}
Then the analytical expression of the parameter vector \(\boldsymbol{\theta}_{L}^{i}\) in the i-th iteration is given by
\begin{equation}
  \boldsymbol{\theta}_{L}^{i} = ((\boldsymbol{Z}^{i})^T \boldsymbol{Z}^{i})^{-1} (\boldsymbol{Z}^{i})^T \boldsymbol{Y}^{i}
\end{equation}
Although batch least squares is adopted in this work, the same data-driven formulation is compatible with recursive parameter-estimation schemes such as RLS, providing a natural basis for future online policy updating.

For the next iteration, the required target policies \(\boldsymbol{u}^{i+1}\) and \(\boldsymbol{w}^{i+1}\) are derived from \(\boldsymbol{\theta}_{L}^{i}\). Once the norm difference of the parameter vector or the TD residual satisfies the preset threshold condition in a certain iteration, the optimal parameter \(\boldsymbol{\theta}_{L}^{*}\) is obtained, from which the final \(H_{\infty}\) controller can be acquired.
Although the convergence of this policy has been proven in Reference [18] and the corresponding algorithm has been applied in other fields, several special conditions should be taken into account when it is implemented for motor control scenarios. Such as the selection of the initial behavioral policy, the form of excitation noise, the volume of sample data, and the appropriate parameter estimation method for practical deployment. The relevant implementation settings are specified and evaluated in the subsequent simulation section.

\section{Simulation Results and Discussion}
In this section, the proposed data-driven model-free control algorithm for fundamental current will be verified through simulations. Namely a control framework combining ESO-based disturbance feedforward compensation and IRL-based \(H_\infty\) feedback control.
\subsection{Simulation Setup and Benchmark Controllers}
The simulations are performed on a three-phase permanent magnet synchronous motor (PMSM), and the motor parameters adopted in the simulation are listed in Table 1.
\begin{table}[htbp]
  \centering
  \caption{Motor Parameters for Simulation}
  \begin{tabular}{lcr}
    \toprule
    Motor Parameter & Value & Unit \\
    \midrule
    Stator resistance \(R_s\) & 0.95 & \(\Omega\) \\[8pt]
    Stator inductance \(L_s\) & 8.5 & mH \\[8pt]
    Stator flux linkage \(\psi_f\) & 0.17 & Wb \\[8pt]
    Pole pairs \(p_n\) & 4 & -- \\ [8pt]
    Moment of inertia \(J\) & 0.002 & \(\mathrm{kg\cdot m^2}\) \\[8pt]
    Rated current \(I_N\) & 20 & A \\[8pt]
    Rated speed \(n_N\) & 3000 & rpm \\
    \bottomrule
  \end{tabular}
  \label{tab:sim_motor_param}
\end{table}
To fully demonstrate the effectiveness of the proposed algorithm, comparative experiments are conducted with other advanced current control schemes, including deadbeat predictive current control [1], state-feedback \(H_\infty\) control [25], and model-free predictive current control [9].
First, a brief introduction to these control strategies is provided. Deadbeat control and \(H_\infty\) control are model-based control algorithms. The former is a feedforward control scheme that achieves extremely fast response when the parameters of the controlled plant are sufficiently accurate; the latter belongs to feedback control, which can solve a comprehensive optimal state feedback controller using model information and possesses favorable robustness.

\subsection{Verification of ESO-Based Disturbance Feedforward}
This section first performs open-loop tests for the aforementioned lumped disturbance observer in a PMSM simulation platform with double closed-loop PI control. By observing the tracking performance between the current state estimates provided by the observer and the actual currents of the system, it can be examined whether the ESO can accurately estimate the lumped disturbances during motor operation. Upon finishing open-loop tests, the feedforward compensation law is applied to the system input, and the resulting closed-loop responses are evaluated.

The motor starts with no load and accelerates to 1000 rpm. A load torque of 10 \(N\cdot m\) is applied at 0.1 s, and the motor further accelerates to 1500 rpm at 0.2 s. The waveforms of speed, torque and d-q axis currents during this process are shown in Figure 4.
\begin{figure}[htbp]
  \centering
  \includegraphics[width = 0.58\textwidth]{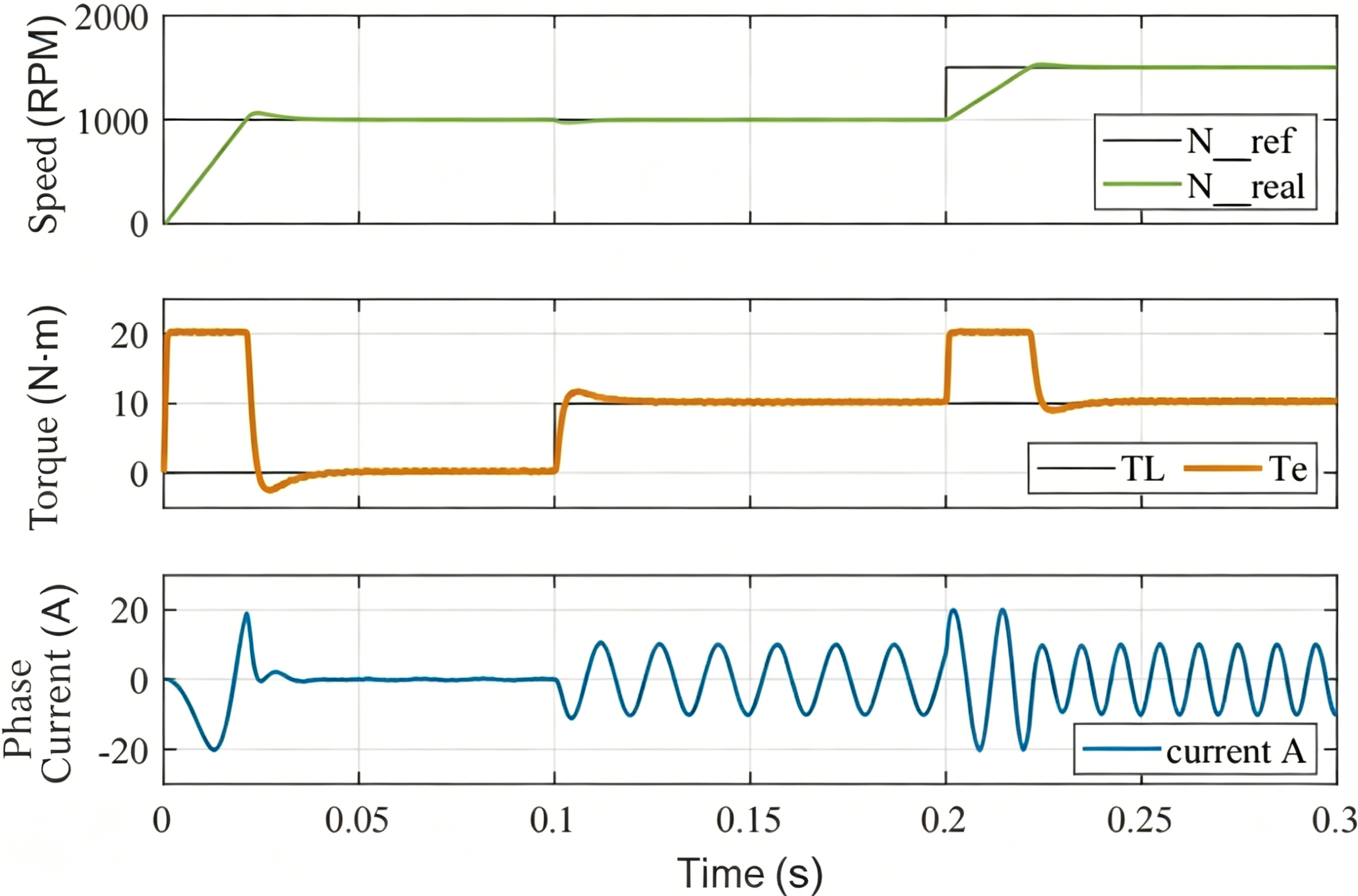}
  \caption{PMSM operating waveforms for ESO validation}
  \label{fig:fig4}
\end{figure}

The bandwidth of the ESO is set to 5000, and \(\alpha_\mathrm{s}\) is taken as the reciprocal of the inductance parameter. The comparison between the actual d-q axis currents and their estimates obtained by the observer is illustrated in Figure 5. As can be seen, the observer realizes zero-error tracking for d-q axis currents. The estimates lead the actual currents by one sampling step in transients, since the observer provides predictions for the subsequent instant.
\begin{figure}[htbp]
  \centering
  \includegraphics[width = 0.58\textwidth]{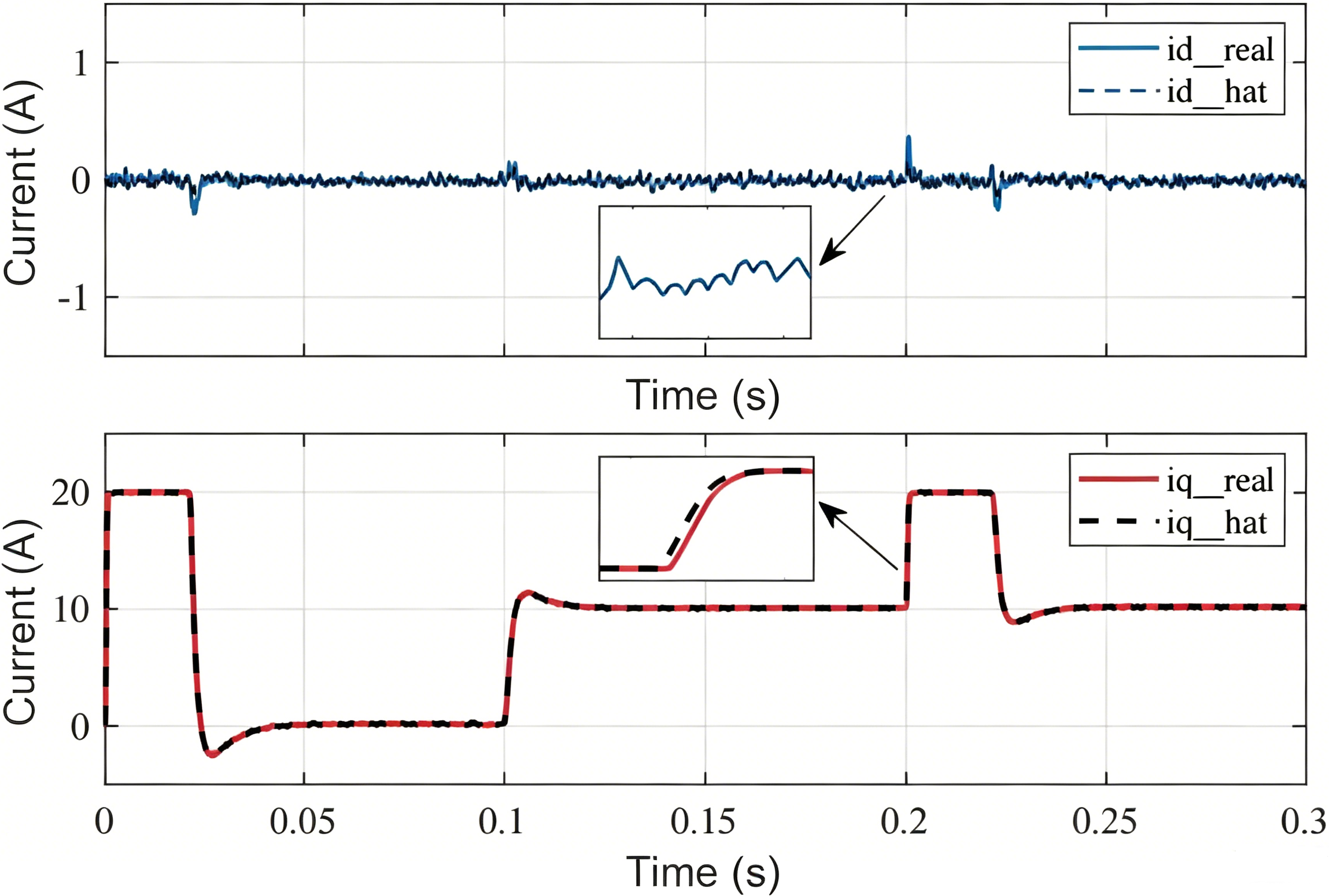}
  \caption{ESO estimation of the \(d\)-axis and \(q\)-axis currents}
  \label{fig:fig5}
\end{figure}

In practical applications, the true values of disturbance states are inaccessible, so the performance of disturbance estimation can only be inferred indirectly from the accuracy of current estimation. But the true lumped disturbances can be calculated using ideal motor parameters in simulation, which provides a benchmark to verify the estimation accuracy of the disturbance observer. The comparison between the actual d-q axis disturbances and the observer's estimated disturbances is presented in Figure 6. Relative to the magnitude of the true disturbances shown in the figure, the observer achieves nearly zero steady-state tracking. Nevertheless, the red curves of disturbance estimation errors reveal persistent fluctuations with amplitudes up to hundreds, and such error oscillations become more severe during dynamic transients. This indicates that notable discrepancies still exist between the estimated and actual disturbances even under perfectly matched nominal parameters. This further demonstrates the necessity of designing a feedback controller in the subsequent section to suppress such disturbance estimation errors.
\begin{figure}[htbp]
  \centering
  \includegraphics[width = 0.5\textwidth]{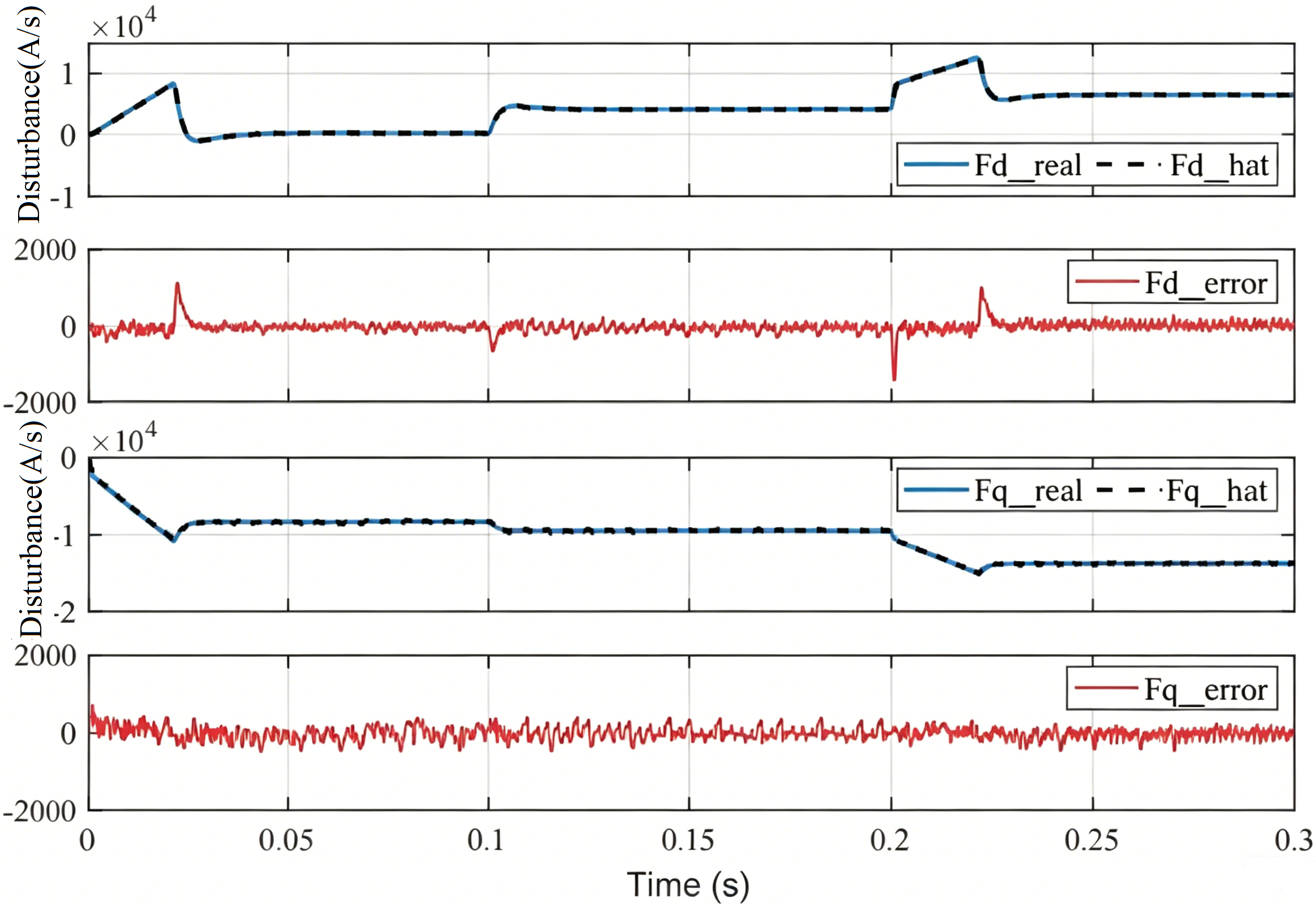}
  \caption{ESO estimation of the \(d\)-axis and \(q\)-axis lumped dynamics under nominal scaling}
  \label{fig:fig6}
\end{figure}
The above disturbance observer design adopts \(\alpha_\mathrm{s}\), which equals the reciprocal of the inductance parameter. However, in the proposed model-free formulation, the actual motor inductance is not assumed to be accurately known. So, the algorithm is expected to adapt autonomously to time-varying parameters during operation. Hence, it is essential to test the disturbance estimation performance under \(\alpha_\mathrm{s}\) mismatch. The scaling factor \(\alpha_s\) is set to 0.5 and 1.5 times its nominal value. The d-q axis disturbance estimation results under these two mismatched cases are plotted in Figure 7.
\begin{figure}[htbp]
  \centering
  \begin{subfigure}{0.48\textwidth}
    \centering
    \includegraphics[width=\linewidth]{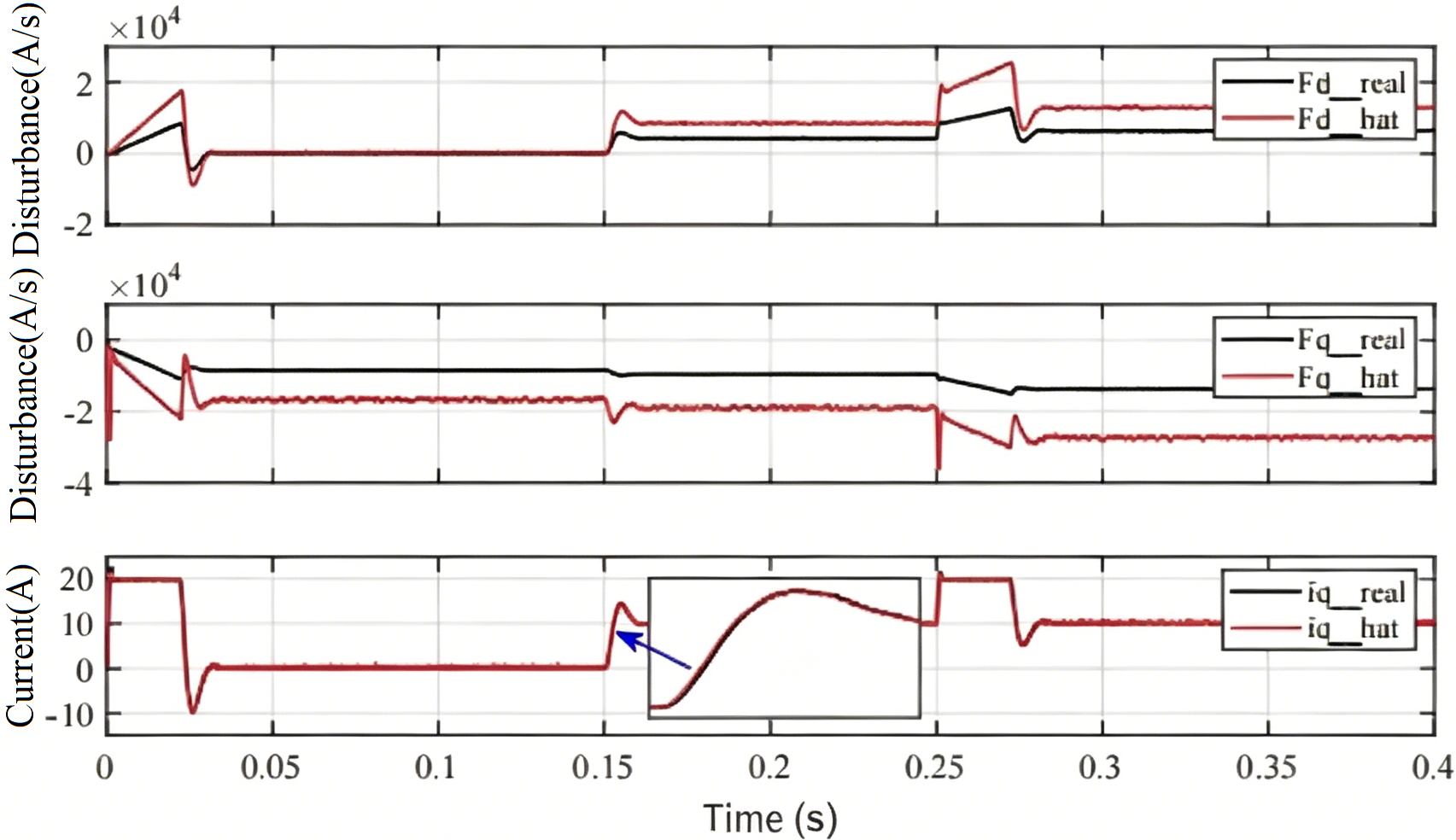}
    \caption{ \(\alpha_s=0.5\alpha_{s,\mathrm{nom}}\)}
  \end{subfigure}
  \hfill
  \begin{subfigure}{0.48\textwidth}
    \centering
    \includegraphics[width=\linewidth]{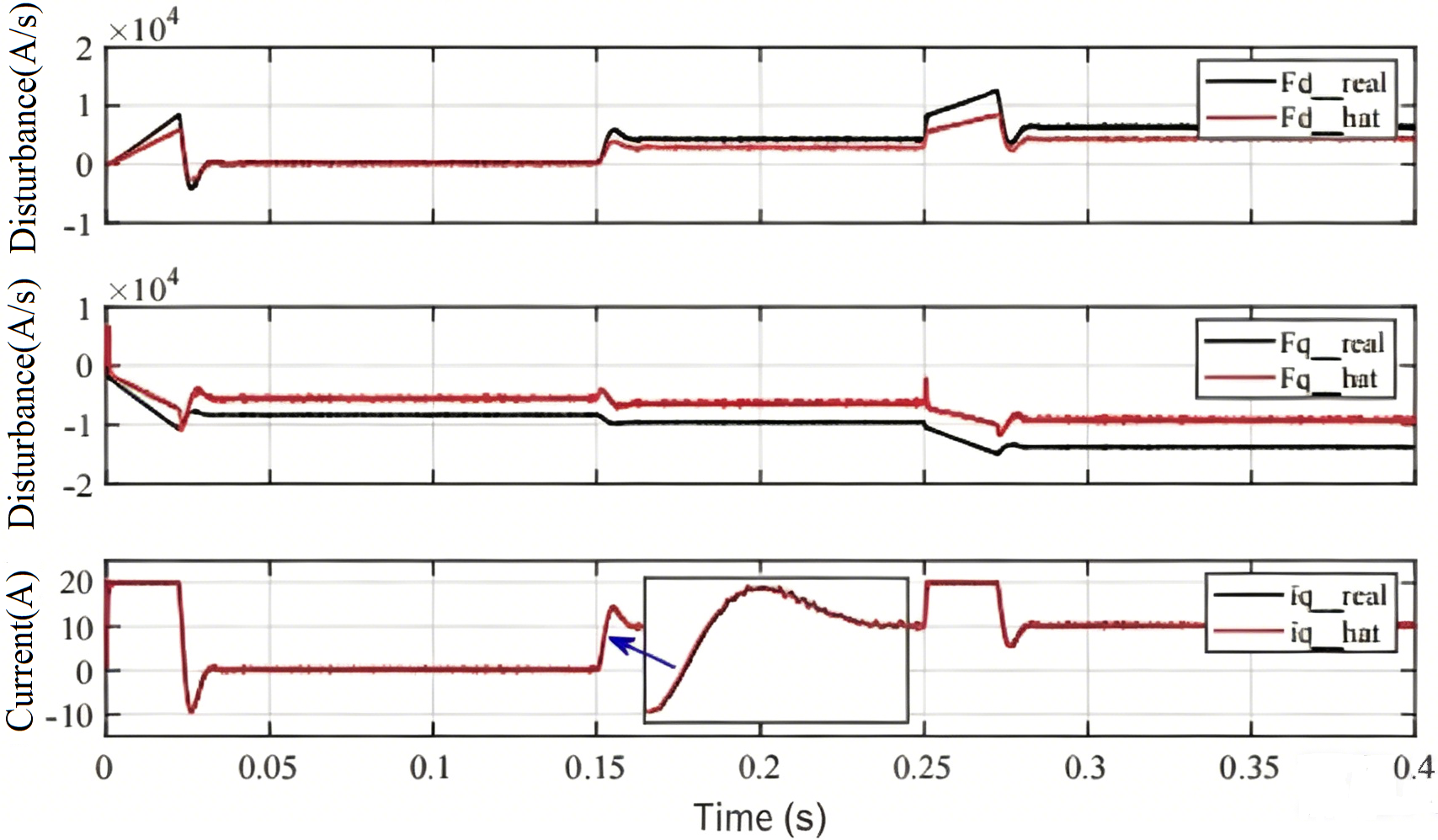}
    \subcaption{\(\alpha_s=1.5\alpha_{s,\mathrm{nom}}\)}
  \end{subfigure}
  \caption{ESO estimation under scaling-factor mismatch}
  \label{fig:7}
\end{figure}

As shown in Figure 7, the observer can still precisely estimate the currents, yet a noticeable steady-state bias emerges between the estimated and real disturbances. Based on the steady-state ESO block diagram in Figure 1,it can be noticed that the estimated disturbance is proportional to the real one, and the proportional gain is exactly the ratio of the mismatched \(\alpha_\mathrm{s}\) to its true value. Further analysis with the feedforward control law \(\hat{F}/\alpha_\mathrm{s}\) reveals that this mismatch-induced proportional gain is completely canceled out in the final feedforward output \(u^\mathrm{ff}\). Consequently, both inductance mismatch and arbitrary selection of \(\alpha_\mathrm{s}\) within an admissible range do not affect the steady-state feedforward compensation value, although they may alter the estimated lumped disturbance and transient estimation behavior.

\subsection{Learning and Convergence of the Residual \(H_\infty \) Controller}

The data-driven algorithm proposed in this paper belong to model-free control, yet it adopts a novel control architecture with theoretically optimal performance. The permanent magnet synchronous motor (PMSM) drive system adopting the proposed control architecture is illustrated in Figure 8.

\vspace*{23pt}
\begin{figure}[htbp]
  \centering
  \includegraphics[width = 0.68\textwidth]{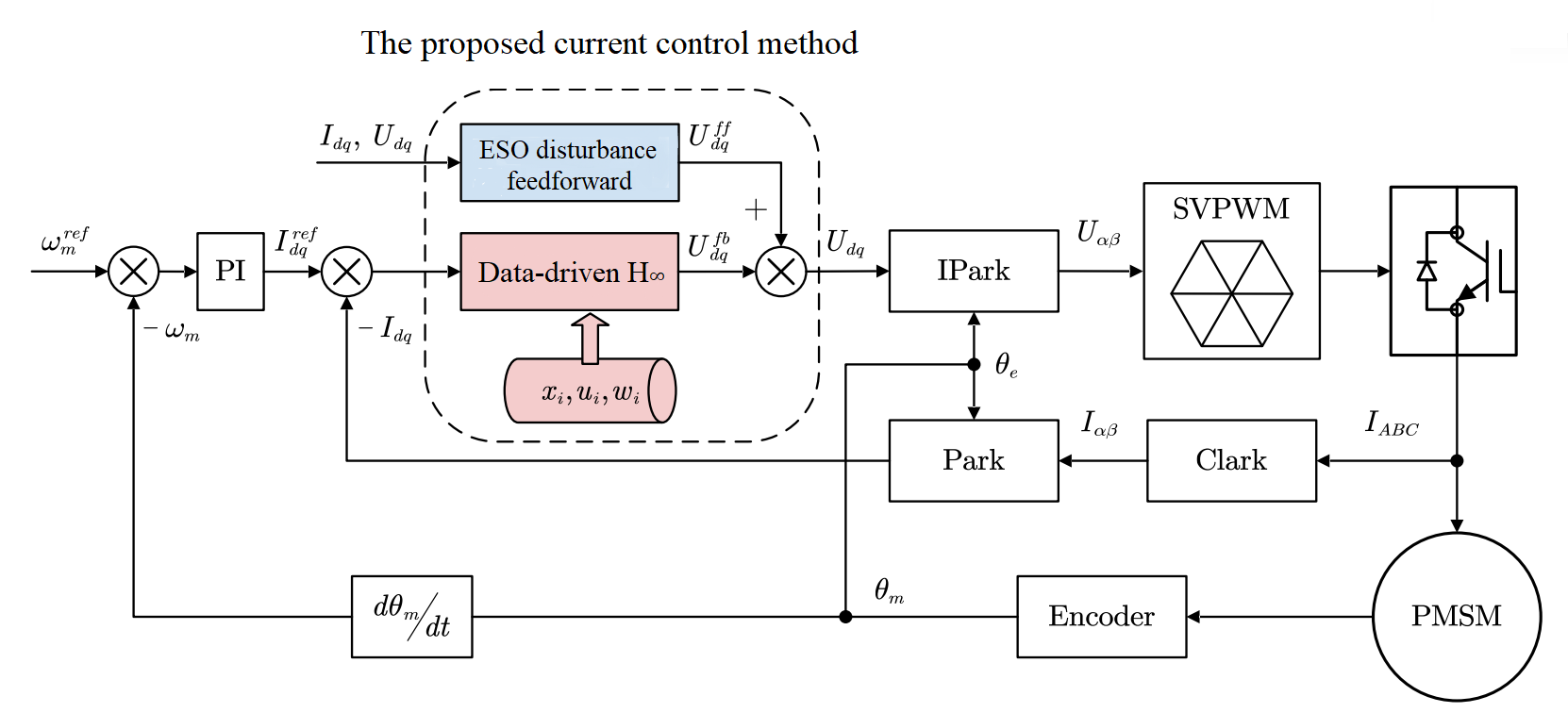}
  \caption{Overall block diagram of the proposed PMSM current-control method}
  \label{fig:fig8}
\end{figure}
\vspace{10pt}
 For an unknown PMSM plant, conventional methods require repeated trial-and-error parameter tuning to achieve favorable current control performance. The proposed data-driven method also needs an initial control policy for data collection, it differs fundamentally in that the data used to train and solve the \(H_\infty\) controller can exhibit poor transient and steady-state performance. This indicates that only simple tuning is required to guarantee system stability. Furthermore, the proposed method relies on model-free disturbance feedforward, which compensates for most system dynamics in advance. In practical applications, a stabilizing admissible behavior policy is employed for data collection. Under this behavior policy, the motor starts with a 5 \(N\cdot m \) load and accelerates to 1000 rpm. Both the current control frequency and inverter switching frequency are set to 10 kHz, and the DC bus voltage is 400 V. The simulation results under this operating condition are presented in Figure 9, which contains the waveforms of d-q axis currents and lumped d-q axis disturbances.

 \vspace{13pt}
 \begin{figure}[htb]
  \centering
  \includegraphics[width = 0.58\textwidth]{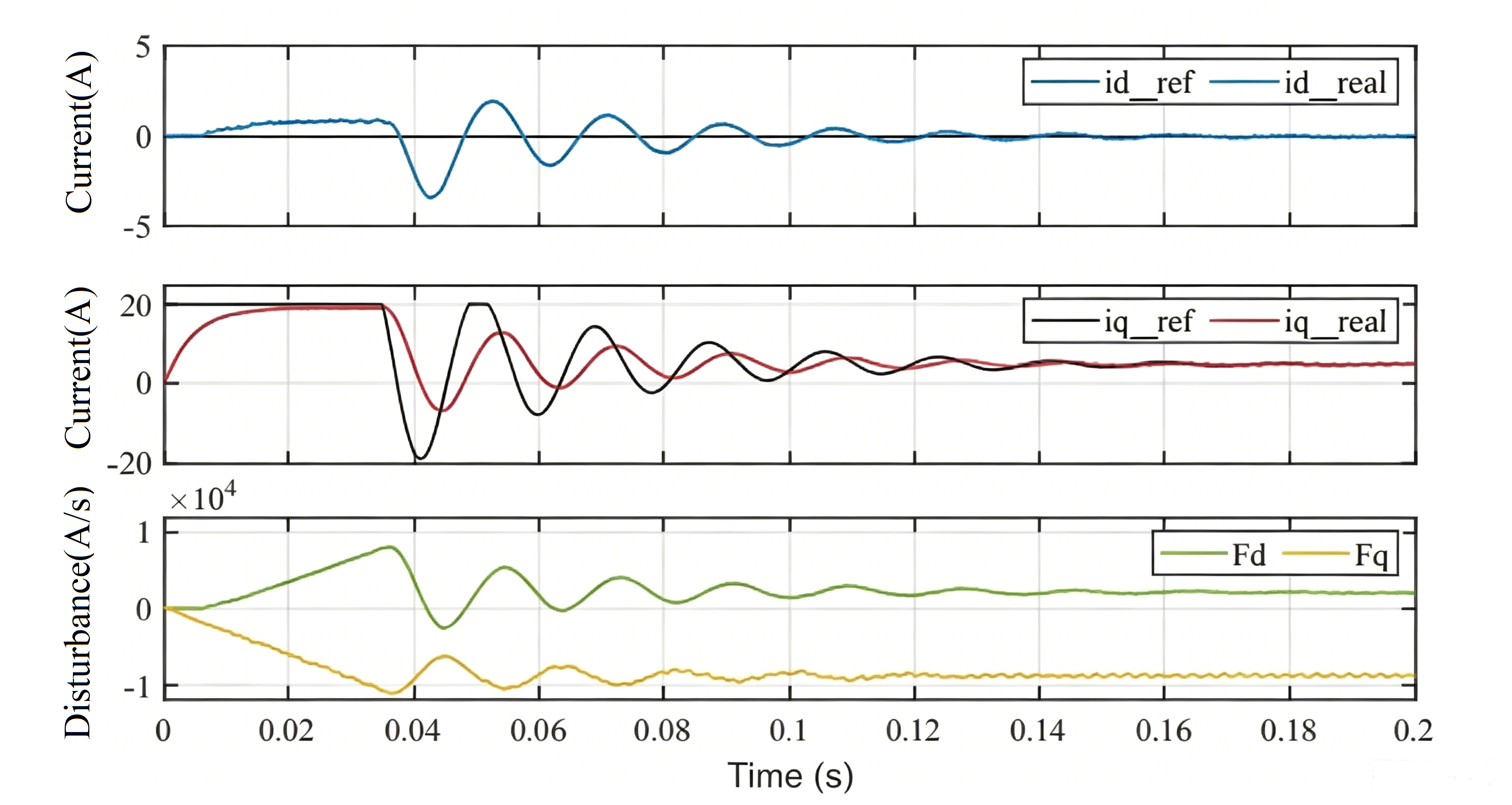}
  \caption{Current and lumped-dynamics trajectories under the data-collection behavior policy}
  \label{fig:fig9}
\end{figure}
Since the system can be approximated as a second-order linear system, a total of 11 weight parameters need to be identified, which means only a small amount of training data is required and the computational complexity remains low. In this paper, the sample number is set to \(N=50\) and the integral time is \(T=0.001\ \mathrm{s}\). 500 groups of d-q axis current states, feedback control inputs and lumped disturbance residuals are collected within an arbitrary 0.05 s interval in the simulation, corresponding to 500 current control cycles. The collected data are fed into the Off-Policy learning strategy proposed above. The iterative convergence curves of the state feedback matrix and the TD residual are presented in Figure 10,respectively.

\begin{figure}[htbp]
  \centering
  \begin{subfigure}{0.48\textwidth}
    \centering
    \includegraphics[width=\linewidth]{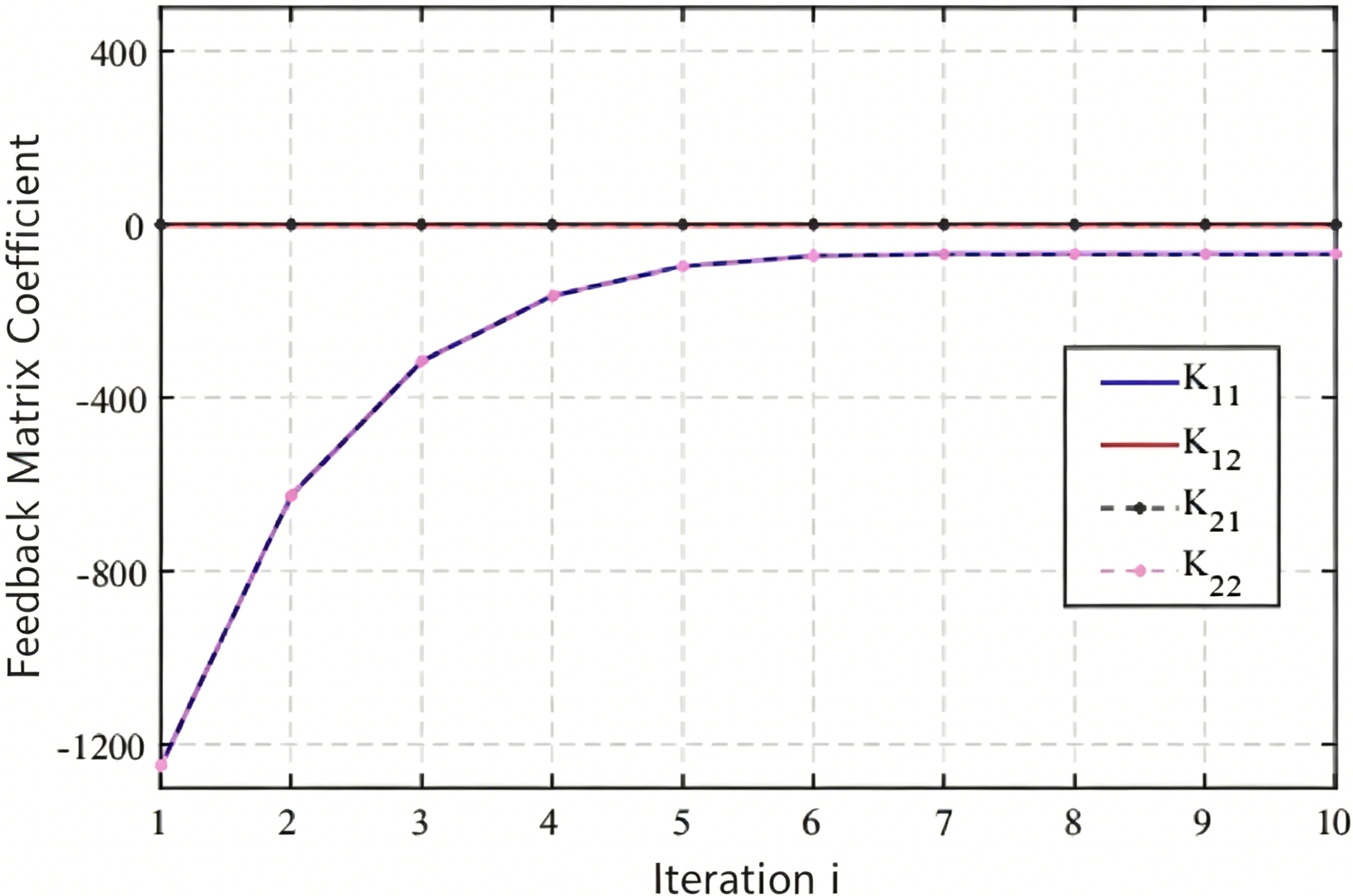}
    \caption{Convergence of the state-feedback gain matrix entries}
  \end{subfigure}
  \hfill
  \begin{subfigure}{0.48\textwidth}
    \centering
    \includegraphics[width=\linewidth]{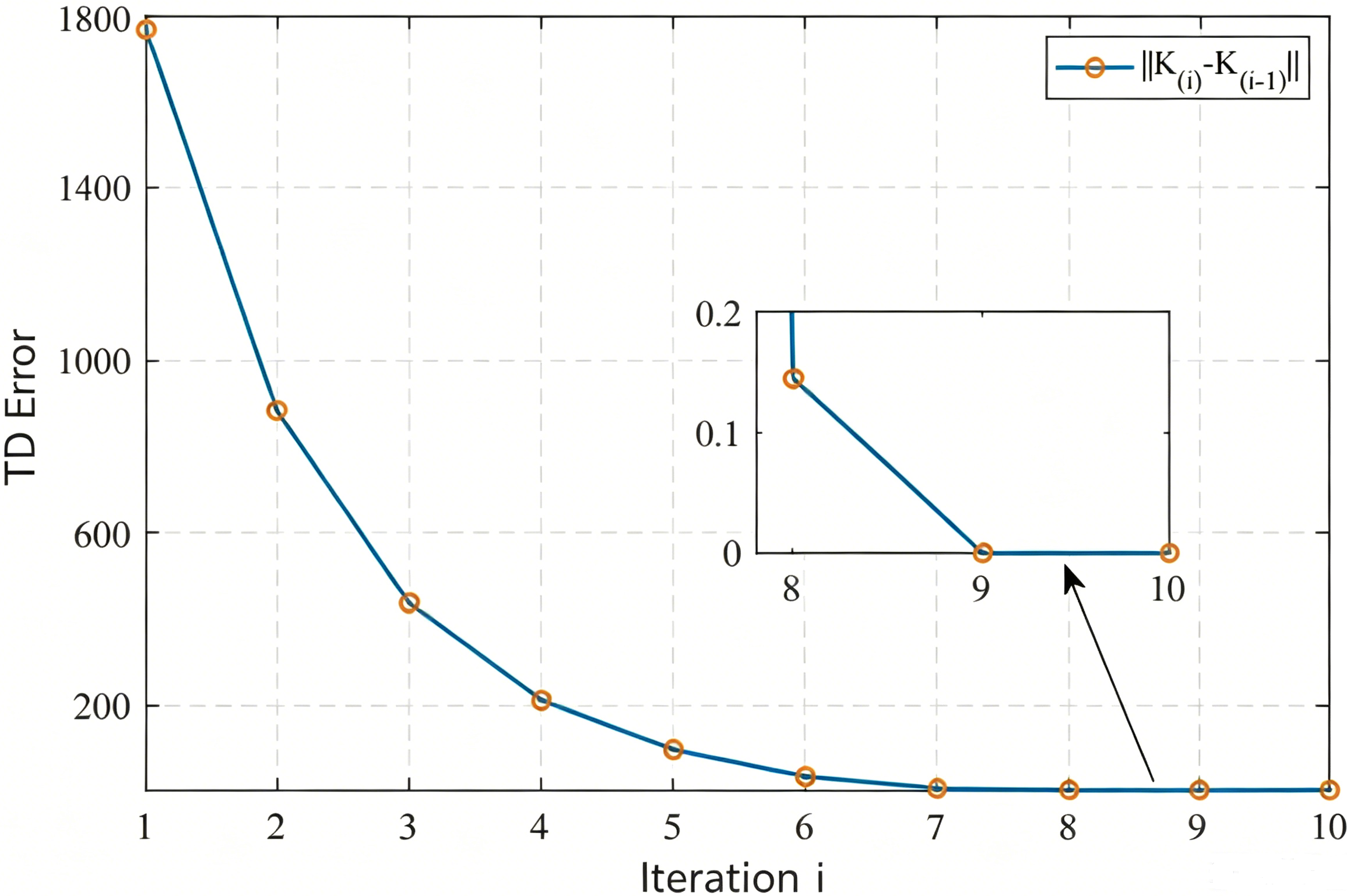}
    \caption{TD residual convergence}
  \end{subfigure}
  \caption{Convergence of the off-policy learning process}
  \label{fig:10}
\end{figure}
\FloatBarrier

At this point, the optimal \(H_\infty\) state feedback matrix is \(\boldsymbol{K}^*=\begin{bmatrix}69.177 & 0 \\ 0 & 69.177\end{bmatrix}\), and the final feedback control law is obtained as \(\boldsymbol{u}_{\text{fb}} = -\boldsymbol{K}^*\boldsymbol{x}\). Subsequently, the complete feedforward-feedback control law is applied to the system, which takes the form \(\boldsymbol{u} = -\hat{\boldsymbol{F}}/\alpha_s + \dot{\boldsymbol{i}}^{ref}/\alpha_s- \boldsymbol{K}^*\boldsymbol{x}\). The corresponding simulation results are shown in Figure 11. It can be observed that the d-q axis currents can stably track the reference signals with ultrafast dynamic response and high steady-state accuracy. Compared with the initial control policy, the control performance is greatly improved, which preliminarily verifies the effectiveness of the proposed current control strategy in this paper.
\begin{figure}[htb]
  \centering
  \includegraphics[width = 0.58\textwidth]{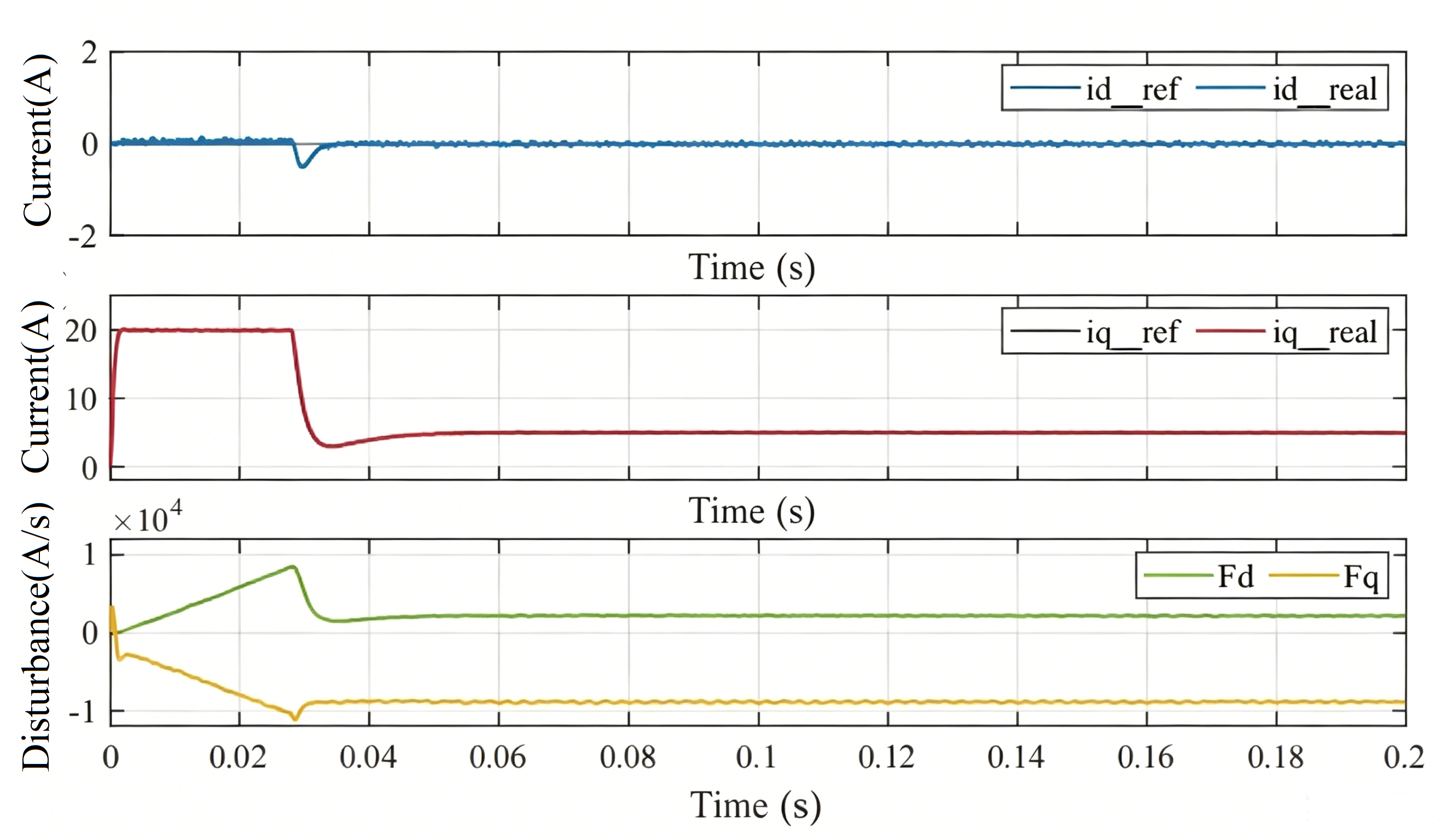}
  \caption{Closed-loop response using the learned \(H_\infty\) residual-feedback policy}
  \label{fig:fig11}
\end{figure}
\subsection{Current-Control Performance under Nominal Parameters}

To quantitatively analyze the performance of the proposed current control strategy, deadbeat predictive current control, state-feedback \(H_\infty\) control and model-free predictive current control are adopted for comparative simulation tests. The q-axis current dynamic response, controller output voltage and total harmonic distortion (THD) of phase current are taken as the main evaluation indicators. The speed loop of all current control algorithms adopts PI control with identical simulation parameters, which ensures the rigor of comparison. In the simulation, the motor starts under no-load condition and accelerates to 1000 rpm, a 10 \(N\cdot m\) load is applied at 0.1 s, and the speed rises to 1500 rpm at 0.2 s. Figure 12 presents the simulation results of deadbeat predictive current control, state-feedback \(H_\infty\) control, model-free predictive current control and the proposed data-driven current control under nominal parameters, respectively. It can be seen that deadbeat control achieves the best tracking performance and the fastest dynamic response, while the other three methods exhibit comparable performance. Figure 13 illustrates that the proposed method achieves the minimum THD of phase current.
\begin{figure}[htb]
  \centering
  \begin{subfigure}[c]{0.48\textwidth}
    \centering
    \includegraphics[width=\linewidth]{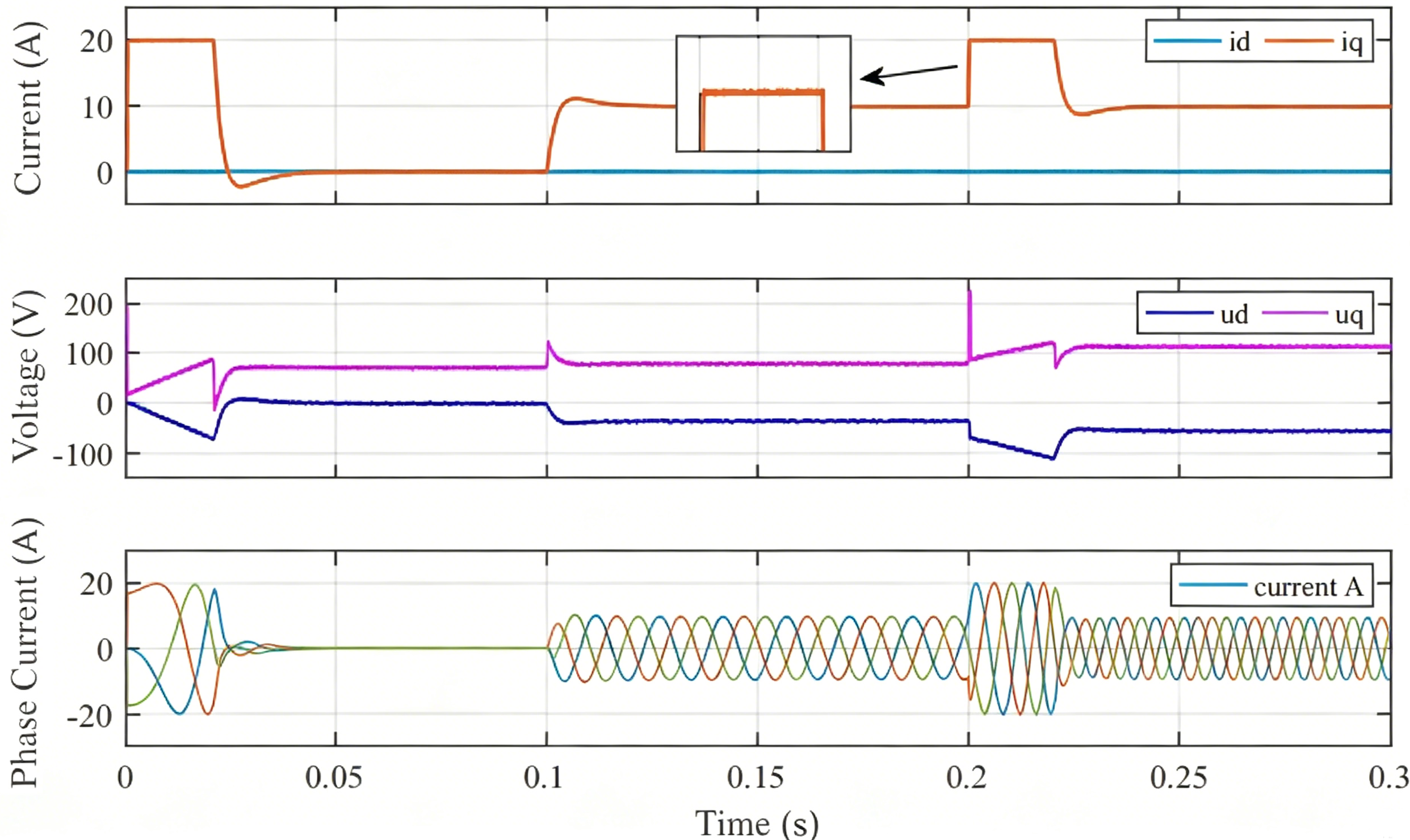}
    \caption{Deadbeat predictive current control}
  \end{subfigure}
  \hfill
  \begin{subfigure}[c]{0.48\textwidth}
    \centering
    \includegraphics[width=\linewidth]{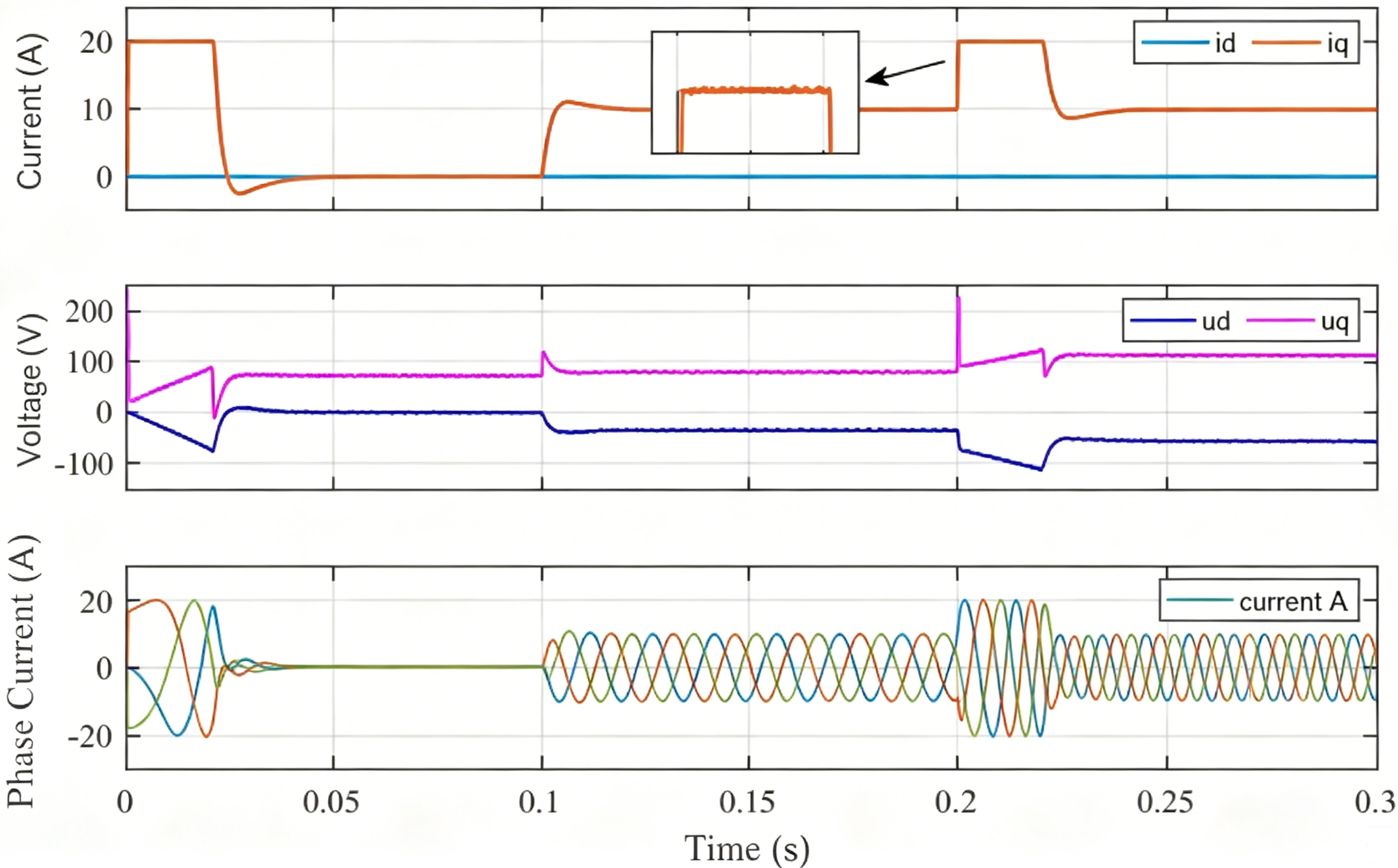}
    \caption{Model-based state-feedback \(H_\infty\) control}
  \end{subfigure}
  \vspace{0.4cm}
  \begin{subfigure}[c]{0.48\textwidth}
    \centering
    \includegraphics[width=\linewidth]{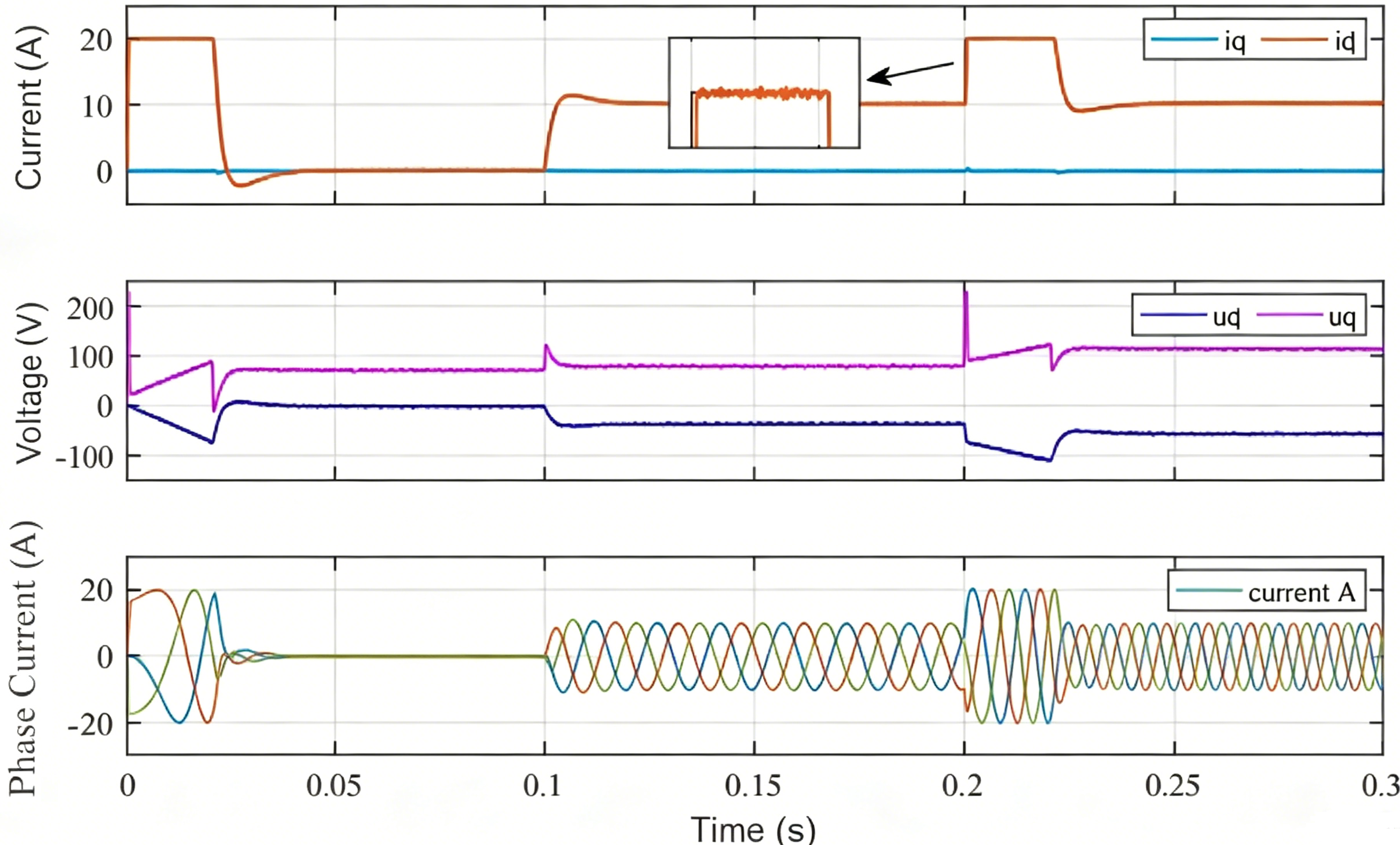}
    \caption{Model-free predictive current control}
  \end{subfigure}
  \hfill
  \begin{subfigure}[c]{0.48\textwidth}
    \centering
    \includegraphics[width=\linewidth]{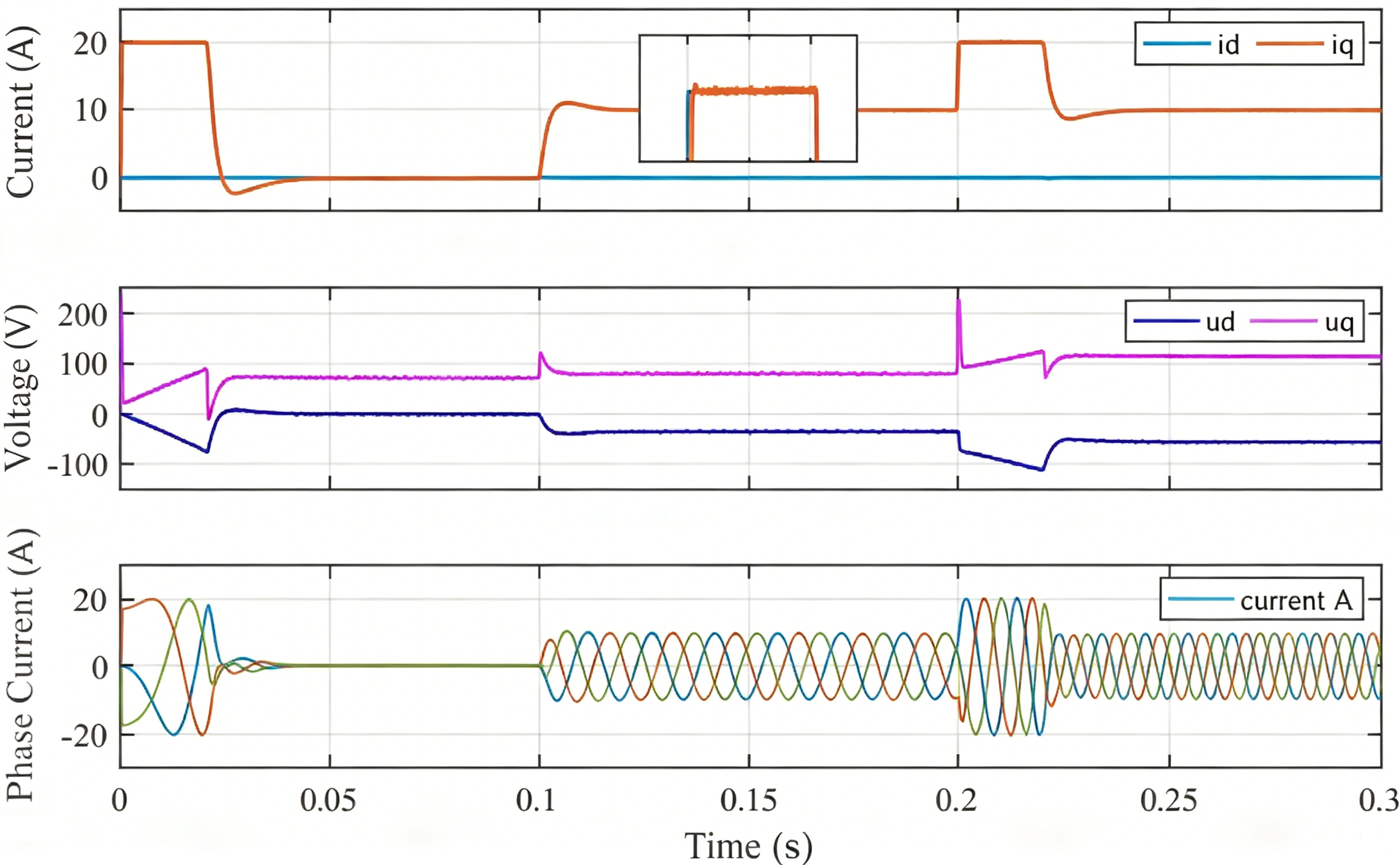}
    \caption{Proposed model-free data-driven current control}
  \end{subfigure}
  \caption{Comparative current-control responses under nominal motor parameters}
  \label{fig:12}
\end{figure}
\begin{figure}[htb]
  \centering
  \includegraphics[width = 0.58\textwidth]{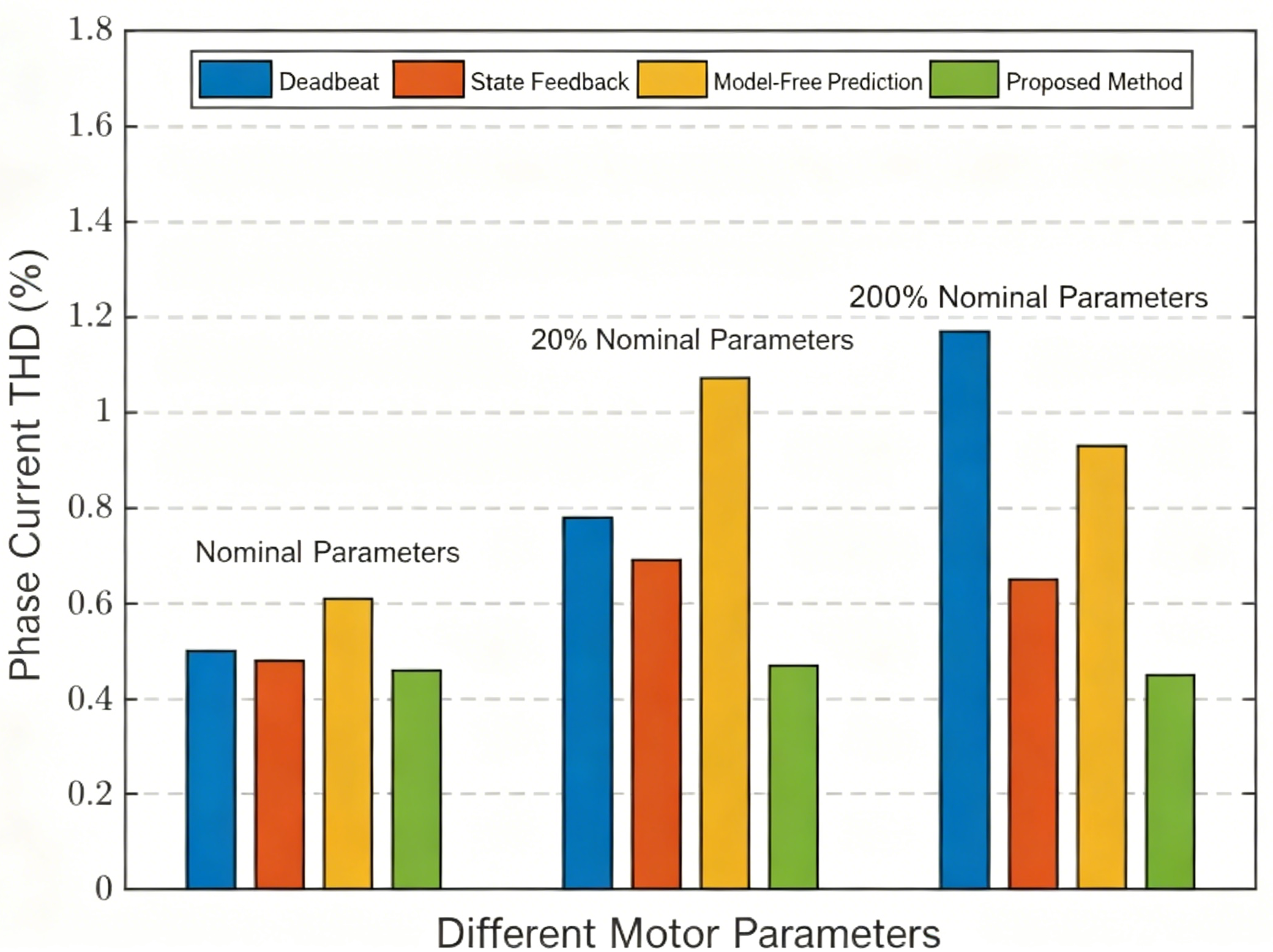}
  \caption{Phase-current THD comparison under nominal and motor-parameter-variation conditions}
  \label{fig:fig13}
\end{figure}
\FloatBarrier

\subsection{Robustness against Motor-Parameter Variations}
To further verify the parameter robustness of the proposed algorithm, the stator inductance, stator resistance and permanent magnet flux linkage are set to 20\% and 200\% of their nominal values, respectively, to compare the operating performance of various current control strategies. The phase current THD results of three groups of comparative experiments are presented in Figure 13. Due to space limitations, only the detailed simulation waveforms under 20\% nominal parameters are provided in Figure 14. Among all schemes, deadbeat predictive current control suffers the most severe impact from parameter mismatch, followed by state-feedback \(H_\infty\) control and model-free predictive current control. In contrast, the proposed data-driven control strategy is barely affected. A detailed analysis of the underlying reasons is given as follows.
\begin{figure}[htbp]
  \centering
  \begin{subfigure}[c]{0.48\textwidth}
    \centering
    \includegraphics[width = \linewidth]{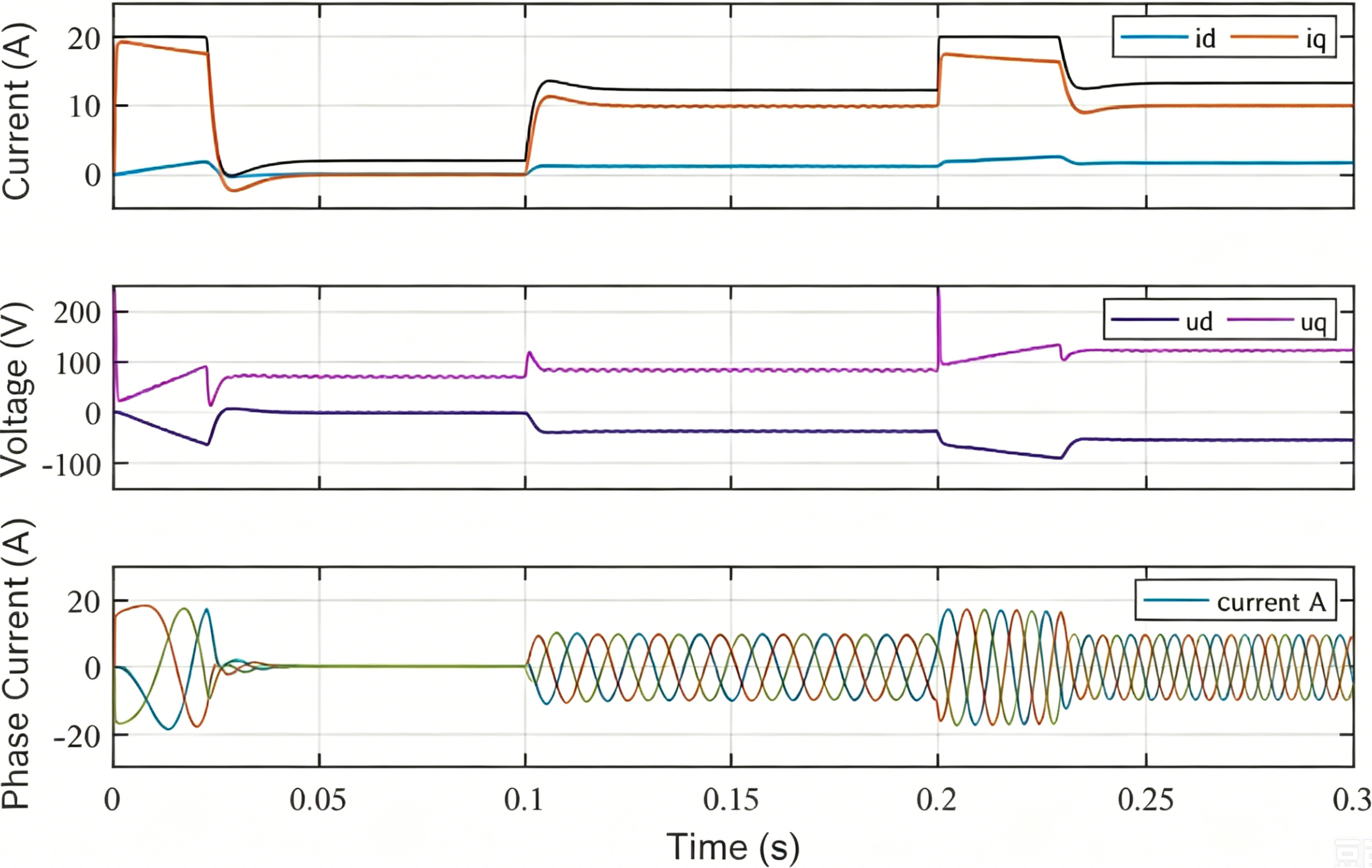}
    \caption{Deadbeat predictive current control}
  \end{subfigure}
  \hfill
  \begin{subfigure}[c]{0.48\textwidth}
    \centering
    \includegraphics[width = \linewidth]{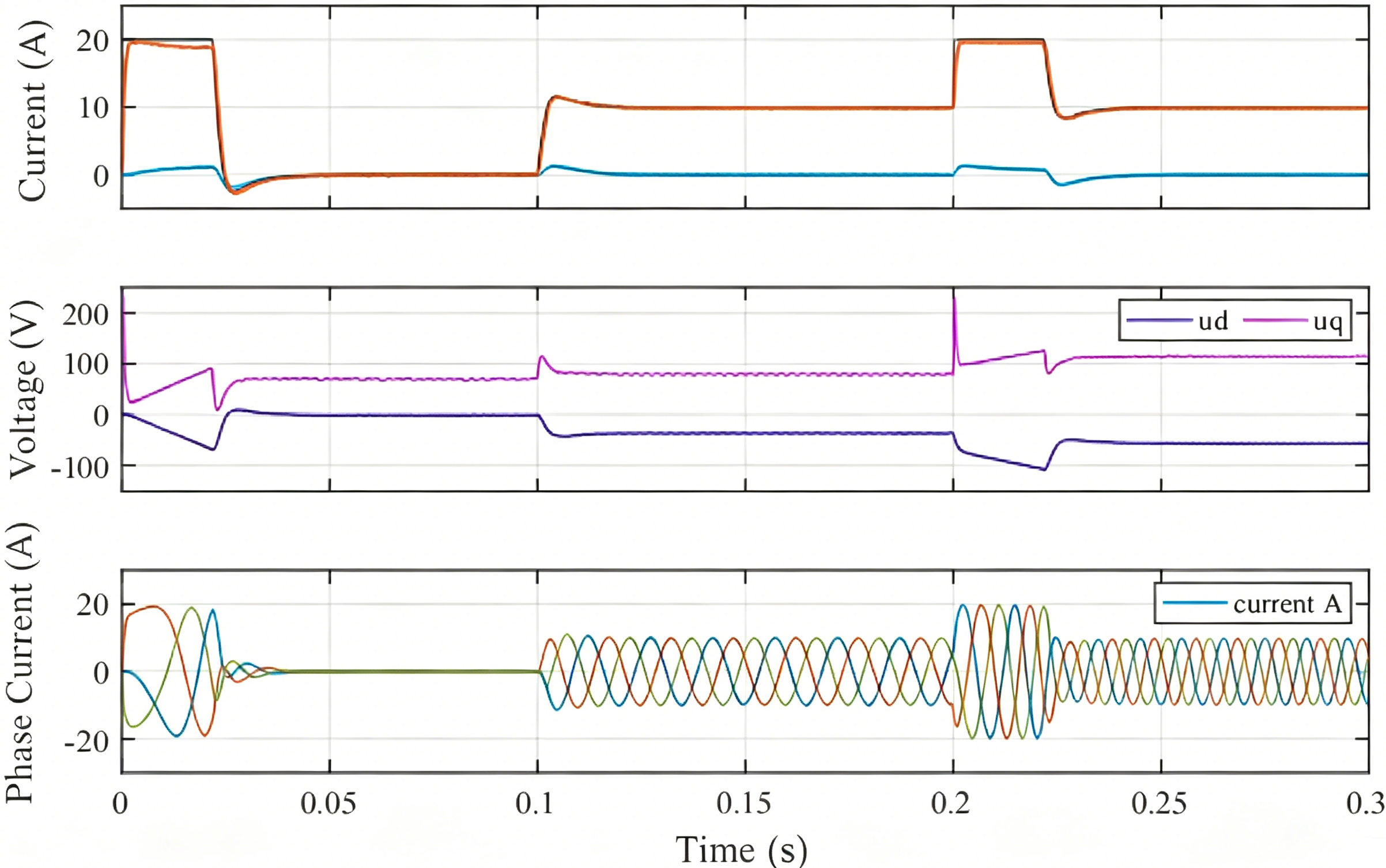}
    \caption{Model-based state-feedback \(H_\infty\) control}
  \end{subfigure}
  \vspace{0.5cm}
  \begin{subfigure}[c]{0.48\textwidth}
    \centering
    \includegraphics[width = \linewidth]{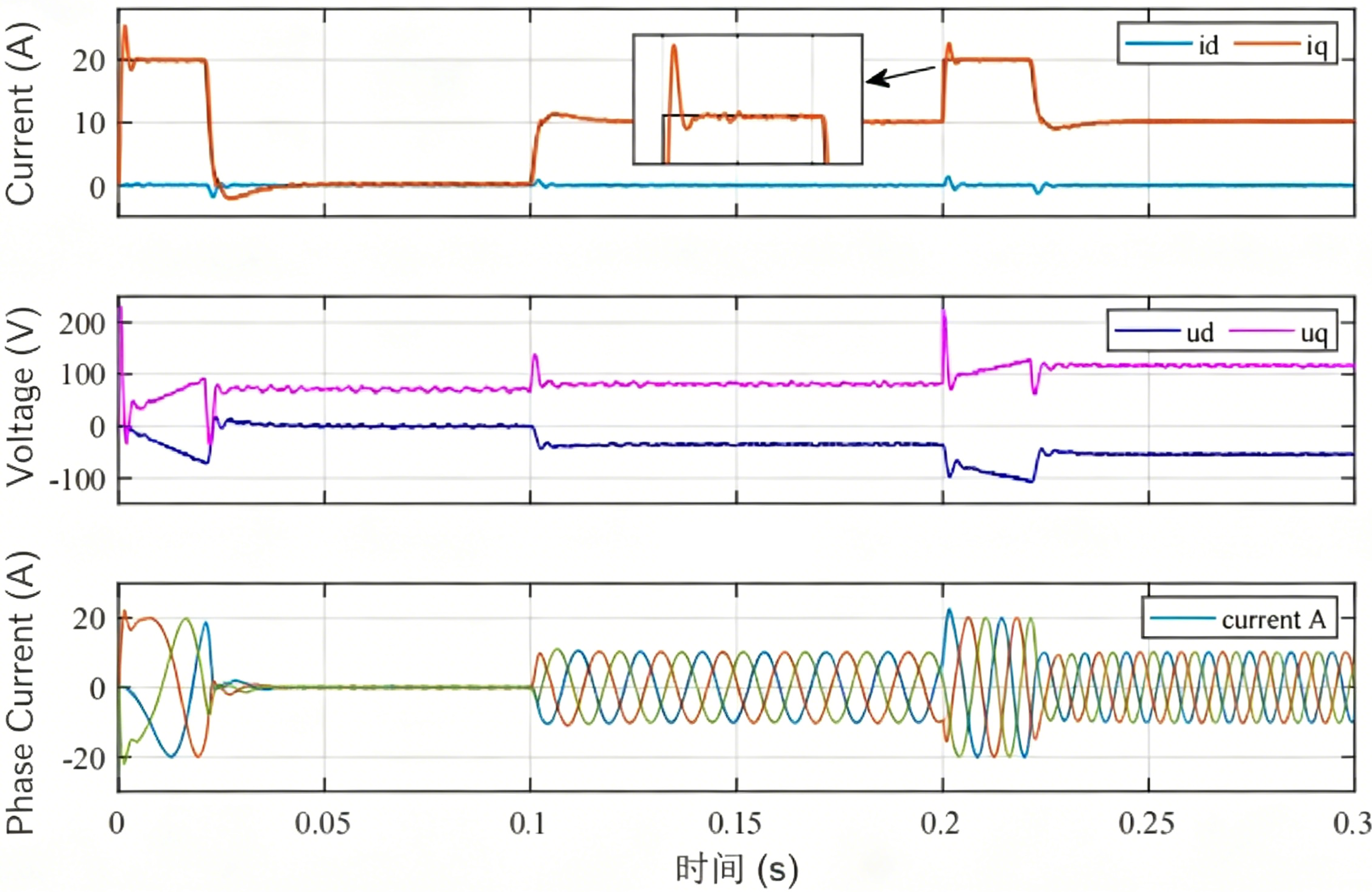}
    \caption{Model-free predictive current control}
  \end{subfigure}
  \hfill
  \begin{subfigure}[c]{0.48\textwidth}
    \centering
    \includegraphics[width = \linewidth]{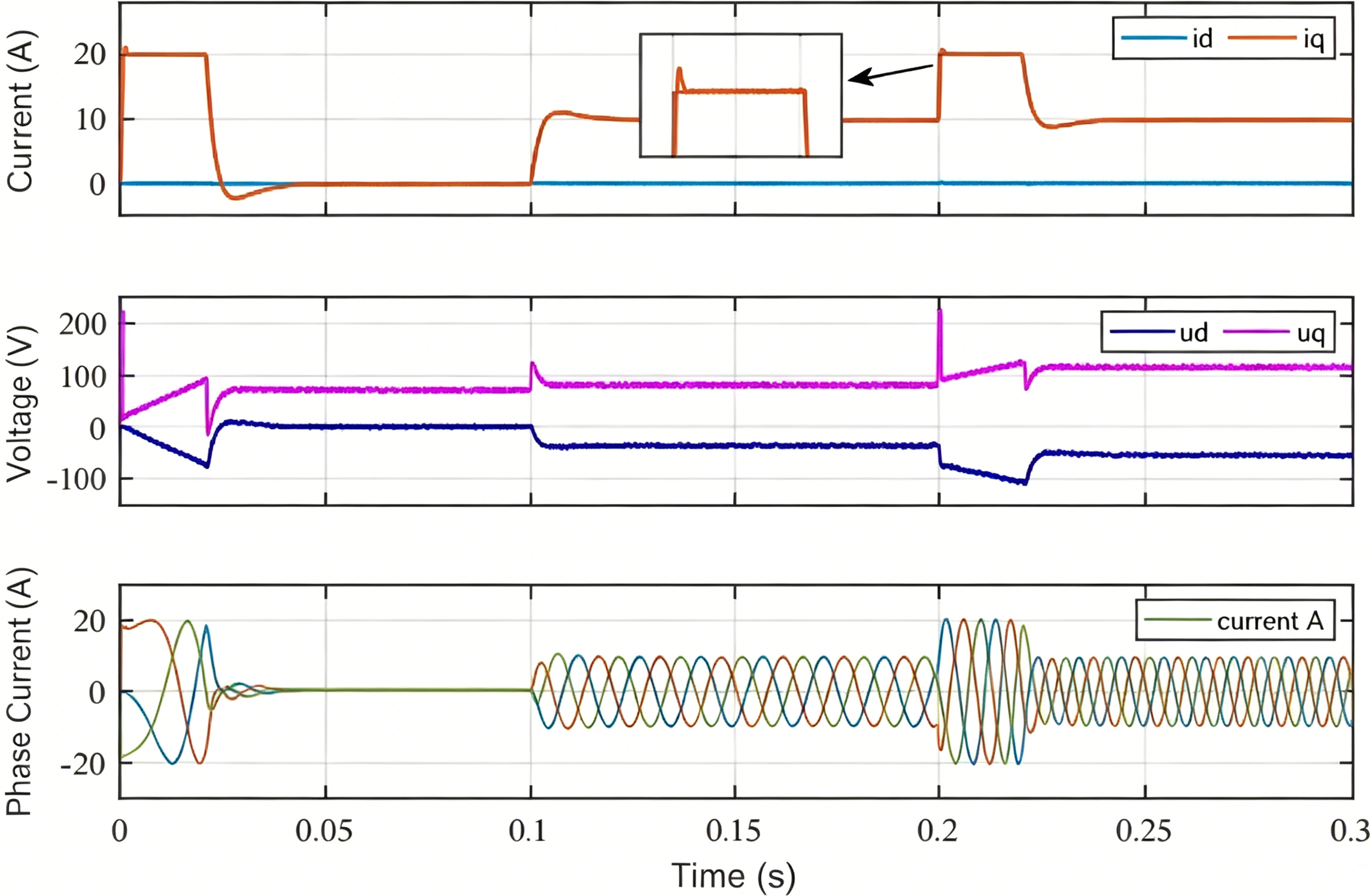}
    \caption{Proposed model-free data-driven current control}
  \end{subfigure}
  \caption{Comparative current-control responses with the motor electrical parameters set to 20\% of their nominal values}
  \label{fig:14}
\end{figure}

Deadbeat control essentially relies entirely on the accurate mathematical model of the current loop to calculate the required reference voltage and it has no capability for feedback correction.
When electrical parameter mismatch occurs, the calculation results derived from the prediction model become inaccurate, thereby generating tracking errors when applied to the actual plant. 
As shown in Figure 14(a), the q-axis current maintains a substantial steady-state deviation with its DC reference throughout the entire process.
State-feedback \(H_\infty \) control solves for the optimal controller based on the state-space model of the controlled plant. As reflected by its objective function, this scheme possesses a certain capacity for disturbance rejection, which yields a lower phase current THD. Nevertheless, its control performance degrades once disturbances exceed the pre-designed robustness constraints. In the simulation, the parameters adopted for controller design deviate significantly from the actual motor parameters. Consequently, As shown in Figure 14(b), the current tracking responses become slower and have steady-state deviations during motor startup and speed acceleration. 
Model-free predictive current control achieves better performance compared with the two foregoing methods, and can realize zero-error tracking even under severe parameter mismatch, since its implementation barely relies on plant model parameters. However, Figure 14(c) reveals prominent overshoots of the q-axis current and obvious fluctuations in d-q axis voltages during startup and acceleration, also resulting in relatively high phase current THD. This indicates that both dynamic and steady-state performance deteriorate to a certain extent under parameter variations. The underlying reason is that, despite its disturbance estimation and feedforward links being model-independent, the predictive control gains within its control law are fixed and cannot guarantee optimal operation in the presence of parameter mismatch.
By contrast, the proposed method combines ESO-based disturbance feedforward compensation with data-driven residual feedback. In all parameter-variation tests, the learned \(H_\infty\) feedback policy is kept unchanged, with no additional data collection, retraining, or controller retuning. The resulting robustness therefore arises from the feedforward-residual-feedback structure and the disturbance-rejection capability of the learned feedback policy, rather than from online controller adaptation.It can be observed from Figure 14(d) that the proposed scheme maintains nearly unchanged performance in steady-state tracking and dynamic response under parameter mismatch, featuring the minimum phase current THD and the best robustness among the compared methods under the tested conditions.

\section{Conclusion}
This paper developed a model-free current-control method for PMSMs based on a functional decomposition into disturbance feedforward and residual feedback. By incorporating the unknown motor dynamics and parameter variations into generalized lumped disturbances, the ESO-based feedforward channel compensates for the dominant plant dynamics, while the remaining post-compensation dynamics are regulated by a data-driven \(H_\infty\) feedback controller. This decomposition allows the feedback-learning problem to be formulated on a simplified residual system rather than on the original PMSM dynamics, leading to a low-dimensional off-policy solution without neural-network approximation.

The analysis further shows that mismatch between the selected ultra-local scaling factor and the physical current-input gain can be incorporated into the ESO-estimated lumped dynamics. Consequently, accurate PMSM electrical parameters are not required for either feedforward synthesis or data-driven residual-feedback design. Comparative simulations demonstrate fast current tracking, low current distortion, and strong robustness to large electrical-parameter variations. Notably, the learned \(H_\infty\) feedback policy remains fixed during the parameter-variation tests, without retraining or motor-specific controller retuning, while maintaining nearly unchanged control performance.

These results demonstrate that separating dominant lumped-dynamics compensation from residual robust regulation provides an effective route toward motor-parameter-independent PMSM current control. Future work will focus on experimental validation and further evaluation of the proposed method under practical drive conditions.

\clearpage
\nocite{*} 
\bibliographystyle{unsrt} 
\bibliography{references}

\end{document}